\documentclass[reprint,superscriptaddress,prx,amsmath,amssymb,aps]{revtex4-2}

\usepackage{graphicx}

\usepackage[normalem]{ulem}
\usepackage{appendix}    
\usepackage{mathtools}
\usepackage{dcolumn}
\usepackage{amsthm}
\usepackage{amsfonts}
\usepackage{ctable}
\usepackage{bm}
\usepackage{bbm}
\usepackage{wasysym}
\usepackage{color}

\usepackage{comment}
\newcommand{\jc}[1]{\textcolor{blue}{#1}}
\newcommand{\srp}[1]{\textcolor{red}{#1}}

\newcommand{\CX}{\mathrm{CX}}
\newcommand{\CZ}{\mathrm{CZ}}

\usepackage{hyperref}
\renewcommand{\jc}[1]{\textcolor{black}{#1}}
\renewcommand{\srp}[1]{\textcolor{black}{#1}}

\begin{document}

\title{Tailored cluster states with high threshold under biased noise}

\author{Jahan Claes}
\affiliation{Department of Applied Physics, Yale University, New Haven, Connecticut 06511, USA}
\affiliation{Yale Quantum Institute, Yale University, New Haven, Connecticut 06511, USA}
\author{J. Eli Bourassa}
\affiliation{Xanadu, Toronto, Ontario, M5G 2C8, Canada}
\author{Shruti Puri}
\affiliation{Department of Applied Physics, Yale University, New Haven, Connecticut 06511, USA}
\affiliation{Yale Quantum Institute, Yale University, New Haven, Connecticut 06511, USA}

\begin{abstract}
Fault-tolerant cluster states form the basis for scalable measurement-based quantum computation. Recently, new stabilizer codes for scalable circuit-based quantum computation have been introduced that have very high thresholds under biased noise where the qubit predominantly suffers from one type of error, e.g. dephasing. However, extending these advances in stabilizer codes to generate high-threshold cluster states for biased noise has been a challenge, as the standard method for foliating stabilizer codes to generate fault-tolerant cluster states does not preserve the noise bias. In this work, we overcome this barrier by introducing a generalization of the cluster state that allows us to foliate stabilizer codes in a bias-preserving way. As an example of our approach, we construct a foliated version of the XZZX code which we call the XZZX cluster state. We demonstrate that under a circuit-level noise model, our XZZX cluster state has a threshold more than double the usual cluster state when dephasing errors are more likely than errors which cause bit flips by a factor of $O(100)$ or more.
\end{abstract}

\date{\today}

\maketitle

\section{Introduction}


\jc{Measurement-based quantum computing (MBQC) is an alternative to the circuit model of quantum computing~\cite{raussendorf2001one}. Depending on the physical primitives available on the underlying quantum hardware, MBQC can provide a more natural framework for composing quantum algorithms, and has been considered for photonic~\cite{nielsen2004optical,browne2005resource,gilchrist2007efficient,menicucci2006universal,bourassa2021blueprint}, trapped-ion ~\cite{lanyon2013measurement,stock2009scalable}, quantum dot ~\cite{guo2007one,lin2008generation,weinstein2005quantum,tanamoto2006producing}, superconducting ~\cite{tanamoto2006producing,you2007efficient,albarran2018one}, and  neutral atom ~\cite{kuznetsova2012cluster} architectures.} In MBQC, computation is performed by preparing an entangled resource state, known as a cluster state, and then performing measurements on the cluster state to realize quantum gates ~\cite{briegel2001persistent,raussendorf2001one,raussendorf2002one,raussendorf2003measurement,nielsen2005fault,briegel2009measurement}. One can make MBQC fault-tolerant using 3D cluster states, which are cluster states whose measurement outcomes obey constraints that can be used to detect errors. These fault-tolerant cluster states are a universal resource for MBQC, as they can be modified to perform arbitrary computations via braiding ~\cite{raussendorf2006fault,raussendorf2007topological}, lattice surgery ~\cite{herr2018lattice}, or other code deformation techniques ~\cite{brown2020universal}. 

\begin{figure}
    \centering
    \includegraphics[width=\columnwidth]{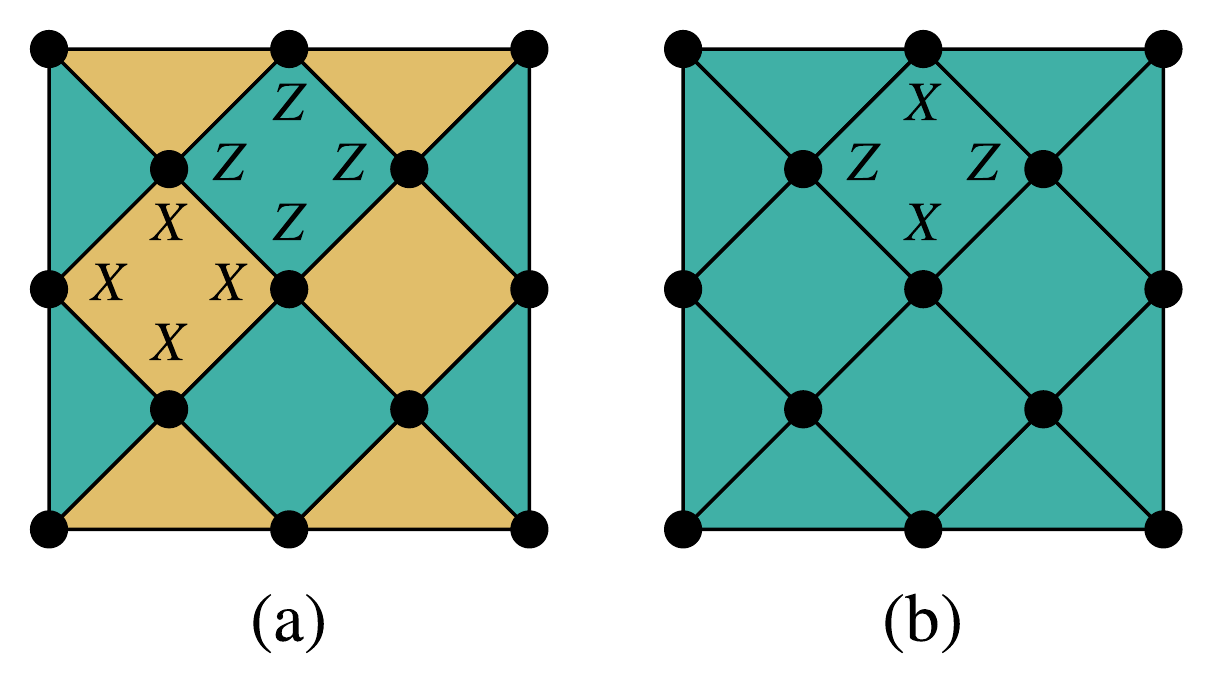}
    \caption{(a) The standard surface code has alternating plaquettes with $X$ and $Z$ stabilizers. (b) The XZZX surface code has the same stabilizer on every plaquette, the product of two $X$ and two $Z$ operators. The XZZX surface code can be obtained from the standard surface code via local Pauli frame changes.}
    \label{fig:SurfaceCodes}
\end{figure}

In circuit model quantum computing, there has recently been intense interest in developing error-correcting codes for qubits with biased noise; an example of biased noise is a scenario in which the \jc{errors which can cause bit-flips are far less likely than those which cause only phase-flips} ~\cite{tuckett2018ultrahigh,tuckett2019tailoring,tuckett2020fault,ataides2021xzzx,darmawan2021practical,higgott2021subsystem}. By modifying the stabilizers of the surface code ~\cite{dennis2002topological} by local Pauli frame change, one can generate the tailored~\cite{tuckett2018ultrahigh,tuckett2019tailoring,tuckett2020fault} or XZZX~\cite{ataides2021xzzx} surface codes (see Fig.~\ref{fig:SurfaceCodes}) as well as other Clifford-deformed surface codes~\cite{dua2022clifforddeformed}, all of which have much higher thresholds than the standard surface code in the presence of biased noise ~\cite{tuckett2018ultrahigh,tuckett2019tailoring,tuckett2020fault,ataides2021xzzx,dua}. The XZZX code is particularly notable for having a high threshold even when using a \jc{simple and efficient} decoder. \jc{Moreover,} bias-preserving controlled-not ($\CX$) gates for stabilizer measurement have also been proposed ~\cite{puri2020bias,guillaud2019repetition,cong2021hardware}, allowing the practical possibility of realizing these surface codes on biased-noise qubits ~\cite{darmawan2021practical}.

However, extending these advances in error-correcting codes to the design of fault-tolerant cluster states for biased-noise qubits remains an open challenge. The standard approach to realize a fault-tolerant cluster state, called \emph{foliation}, is based on performing repeated measurements of the stabilizers of a 2D stabilizer code via teleportation ~\cite{raussendorf2006fault,raussendorf2007topological, bolt2016foliated,brown2020universal}. Indeed, the typical cluster state is created by foliating the standard surface code~\cite{raussendorf2005long,raussendorf2006fault,raussendorf2007topological,raussendorf2007fault}. This cluster state is known as either the Raussendorf-Harrington-Goyal (RHG) or Raussendorf-Bravyi-Harrington cluster state; we will refer to it as the RHG cluster state in this paper. However, we will see below that the process of foliation does not preserve the noise bias, as it effectively converts high-probability pure phase-flip errors into low-probability bit-flip errors. Thus, foliating the tailored or XZZX surface code will not result in a high-threshold cluster state even if the underlying qubits have biased noise. 

In this paper, we introduce a generalized cluster state as a tool for fault-tolerant MBQC. Our generalized cluster state is built by preparing qubits in the $|0\rangle$ and $|+\rangle$ states, and then entangling pairs of qubits using both $\CZ$ and $\CX$ gates. Using our generalized cluster state, we can construct a foliation protocol that preserves the noise bias. This key property allows us to use the generalized cluster state to build high-threshold foliated versions of stabilizer codes designed for biased noise. While our method could be applied to any stabilizer code, we focus on foliating the XZZX surface code as we can decode it efficiently with a \jc{minimum-weight perfect matching} (MWPM) decoder ~\cite{edmonds1965paths,dennis2002topological,kolmogorov2009blossom}. Under biased circuit-level noise, we demonstrate that our new cluster state, the XZZX cluster state, has a much higher threshold than the RHG cluster state. Indeed, the threshold for the XZZX cluster state exceeds $2.2\%$ gate-error rate compared to under $1.0\%$ for the usual RHG cluster state when dephasing errors are more likely than errors that cause bit-flips by a factor of $O(100)$ or more, which is an experimentally realistic amount of noise asymmetry~\cite{puri2020bias,darmawan2021practical}. This work provides a systematic procedure for designing cluster states that are highly effective at correcting structured noise to gain a significant threshold advantage, thus opening up a new direction for realizing near-term practical fault-tolerant quantum technologies ranging from universal MBQC to one-way repeaters for quantum networking and communications \jc{using noisier hardware than would otherwise have been possible}.

\jc{Prior to this work}, there has been recent research into {\it non-foliated} cluster states ~\cite{nickerson2018measurement,newman2020generating}, which cannot be viewed as foliated versions of some 2D stabilizer code. Examples of non-foliated codes with improved thresholds against pure phase-flip errors were found in ~\cite{nickerson2018measurement,newman2020generating}. However, the best-performing of these cluster states require high-degree ($8$-$10$) qubit connectivity and still yield $\sim 12\%$ lower thresholds than what is achieved with the XZZX cluster state presented here which has only degree-4 connectivity. The low qubit connectivity of the XZZX cluster state makes it much more desirable for practical implementation. In addition, ~\cite{stephens2013high} has previously proposed creating high-threshold cluster states by concatenating a base-level repetition code to correct \jc{phase-flip} errors with the usual RHG lattice. This is similar to the approach of ~\cite{xu2019high} to correct biased errors in the circuit-model picture, where they concatenate a repetition code with the usual surface code. The advantage of our approach is similar to the advantage the tailored and XZZX surface codes have over ~\cite{xu2019high}, namely concatenating with a repetition code results in a higher qubit overhead and an enhanced rate of bit-flip errors.

Our paper is structured as follows. First, we review how to construct the RHG cluster state by foliating the surface code. This construction is representative of the more general protocols ~\cite{bolt2016foliated,brown2020universal} for foliating stabilizer codes. Next, we explain why foliation effectively unbiases the qubits, and thus does not give high thresholds when applied to the tailored or XZZX surface codes. We then introduce the generalized cluster state, a simple modification of the usual cluster state that allows us to foliate the XZZX code in a bias-preserving way. We call this the bias-preserving XZZX cluster state. We demonstrate higher thresholds for our XZZX cluster state using a biased circuit-level noise model and minimum weight perfect matching (MWPM) decoder. We conclude with some discussion of experimental platforms that may benefit from our approach. Details of our circuit-level noise model and decoder are presented in the Methods section.

\section{Results}

\subsection{The standard fault-tolerant cluster state}
\label{section:StandardClusterState}
Foliation is a flexible approach to build fault-tolerant cluster states from stabilizer codes~\cite{bolt2016foliated,brown2020universal}; here, we will follow the presentation of folation given in~\cite{brown2020universal}. \jc{The essential idea is to replace each qubit in a stabilizer code with a 1D cluster state that can be used to teleport a single logical degree of freedom, and couple these 1D cluster states so that the stabilizers of the code are repeatedly measured during teleportation.} These cluster states can then be used to fault-tolerantly store some initial encoded state. While the cluster states we construct will not apply logical gates during the teleportation, there are standard methods to modify the basic fault-tolerant cluster state to enable universal fault-tolerant MBQC ~\cite{raussendorf2006fault,raussendorf2007topological, herr2018lattice,brown2020universal,bartolucci2021fusion}.

To illustrate the idea of foliation and motivate our generalized cluster state construction, we will \srp{review} how to build the RHG cluster state by foliating the surface code. \jc{We will begin by constructing the 1D cluster state that can teleport a single qubit. We then replace each surface code qubit with a 1D teleportation cluster state, and demonstrate how to couple them to measure the surface code stabilizers during teleportation. Finally, we will explain how the resulting RHG cluster state detects errors.}

\begin{figure*}
    \centering
    \includegraphics[width=\textwidth]{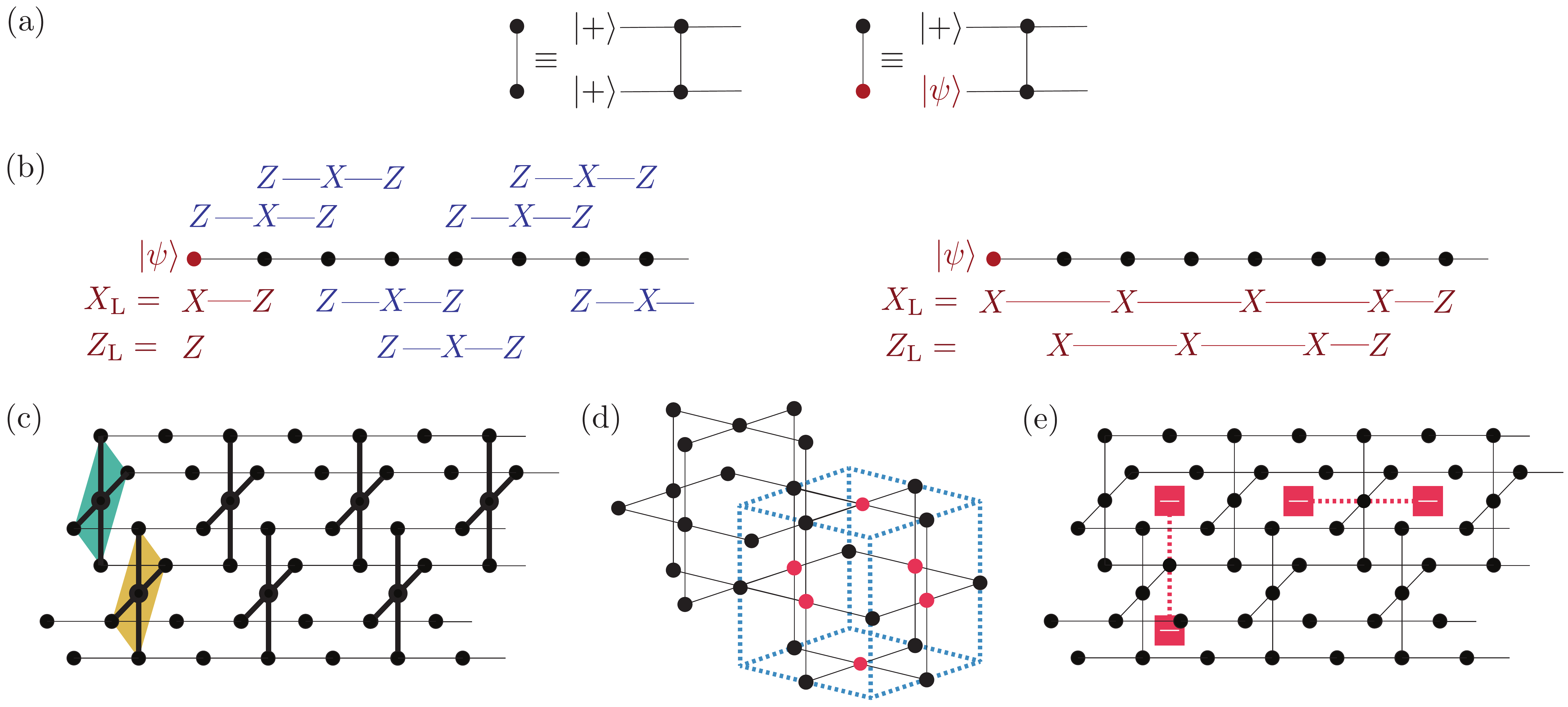}
        \caption{(a) Filled black circles denote qubits initialized in the $|+\rangle$ state, filled red circles denote qubits initialized in an arbitrary state. Qubits connected by a line have a $\CZ$ gate applied between them. (b) Left: The logical operators (red) and stabilizers (blue) of the 1D cluster state. Right: By multiplying the logical operators by stabilizers, we can rewrite them to involve only $X$ operators on the first $2n$ qubits. Measuring the first $2n$ qubits in the $X$ basis teleports logical information to qubits $\{2n+1,2n+2\}$. (c) By coupling ancilla qubits (bold) to a set of 1D cluster states, we can measure multi-qubit logical operators during teleportation. Here, measuring the $X$ operator on the ancilla qubits attached at even (odd) numbered sites measures the logical product of the $Z$ ($X$) operators of the neighboring qubits. We thus measure the stabilizers of the surface code as we teleport. (d) The unit cell of the resulting cluster state, the RHG cluster state. One can verify that the product of $X$ operators on the highlighted qubits is a stabilizer of the RHG cluster state. There is one such stabilizer associated to each cell of the lattice. (e) A $Z$ error on a qubit flips the stabilizers of the neighboring cells, allowing us to detect errors.} 
    \label{fig:OriginalClusterState}
\end{figure*}

The standard 1D teleportation cluster state is illustrated in Figs.~\ref{fig:OriginalClusterState}a,b. To prepare the cluster state, we begin with an arbitrary state $|\psi\rangle$ on the first qubit and a $|+\rangle$ state on the remaining qubits. We then entangle neighboring qubits by applying controlled-phase or $\CZ$ gates to all pairs of neighbors \jc{(Fig.~\ref{fig:OriginalClusterState}a)}. Importantly, the $\CZ$ gates are mutually commuting and may be applied in any order.

The stabilizer formalism can now be used to describe the cluster state. Before the $\CZ$ gates, the logical Pauli operators on $|\psi\rangle$ are given by $X_{\mathrm{L}}=X_1$ and $Z_{\mathrm{L}}=Z_1$, and the state is stabilized by $\{X_2,X_3,\dots\}$. \jc{After applying the $\CZ$ gates, the new logical operators and stabilizers are obtained from the old by conjugating with the $\CZ$ gates. This sends
\begin{equation}
    Z_i\rightarrow Z_i\qquad X_i\rightarrow X_i\prod_{j\in\mathcal{N}_i}Z_j
\end{equation}
where $\mathcal{N}_i$ denotes the neighbors of site $i$. We thus have $X_{\mathrm{L}}=X_1Z_2$ and $Z_{\mathrm{L}}=Z_1$, with stabilizers $\{Z_1X_2Z_3,Z_2X_3Z_4,\dots\}$} (Fig.~\ref{fig:OriginalClusterState}b). By multiplying the logical operators by stabilizers, we can put them in a form involving only $X$ operators on the first $2n$ qubits:
\begin{align}
    X_{\mathrm{L}}&=\left(\prod_{\substack{i\leq 2n\\i\text{ odd}}} X_i\right)X_{2n+1}Z_{2n+2}\label{eq:XLRewriting}\\
    Z_{\mathrm{L}}&=\left(\prod_{\substack{i\leq 2n\\i\text{ even}}} X_i\right)Z_{2n+1}\label{eq:ZLRewriting}
\end{align}
The case of \jc{$n=3$} is illustrated in Fig.~\ref{fig:OriginalClusterState}b. This form makes it clear that if we measure the first $2n$ qubits in the $X$ basis, the logical operators will be teleported to qubits $\{2n+1, 2n+2\}$. For example, in the case \jc{$n=3$}, if we measure qubits \jc{$1-6$} and get outcomes $\{x_1,x_2,x_3,x_4,x_5,x_6\}$ \jc{with $x_i=\pm 1$}, Eqs.~\ref{eq:XLRewriting} and \ref{eq:ZLRewriting} imply that the logical operators are given by
\begin{equation}
        X_{\mathrm{L}} =x_1x_3x_5X_7Z_8,\qquad     Z_{\mathrm{L}} = x_2x_4x_6Z_7.\label{eq:NewXandZLogicals}
\end{equation}

To construct a fault-tolerant cluster state, we combine individual 1D teleportation cluster states, and modify the cluster to measure the check operators of some stabilizer code as we teleport. This construction was first realized for the surface code ~\cite{raussendorf2006fault,raussendorf2007topological,raussendorf2007fault}, but has been adapted to arbitrary CSS codes ~\cite{bolt2016foliated} and non-CSS codes ~\cite{brown2020universal}. To generate the surface code, \jc{we replace each qubit of the surface code shown in Fig.~\ref{fig:SurfaceCodes}a with a 1D teleportation chain}. For each $X$-stabilizer plaquette of the surface code, we attach an ancilla qubit to each set of odd sites of the teleportation chain, and for each $Z$-stabilizer plaquette we attach an ancilla to each set of even sites of the teleportation chain, as shown in Fig.~\ref{fig:OriginalClusterState}c. These ancillas are also initialized in $|+\rangle$ and entangled with their neighbors with $\CZ$ gates. One can easily verify that measuring an ancilla qubit in the $X$ basis \jc{results in} measuring the surface code stabilizer of the corresponding plaquette during the teleportation.

The resulting 3D cluster state, the RHG cluster state ~\cite{raussendorf2006fault,raussendorf2007topological,raussendorf2005long}, is illustrated in Fig.~\ref{fig:OriginalClusterState}d, where we \jc{display} two overlapping cells of the cluster. To fault-tolerantly teleport information through this state, we simply measure both the data qubits and ancilla qubits in the $X$ basis. One can check that for an RHG cluster state without errors, the product of the $X$ operators on the faces of a cell (the highlighted qubits in Fig.~\ref{fig:OriginalClusterState}d) is a stabilizer of the state, so the product of the $X$ measurements around the faces of a cell should be $(+1)$. \jc{To see the effects of errors, we can consider applying a single Pauli error.} A $Z$ or $Y$ error on a face qubit flips the syndromes of the two neighboring cells, allowing us to detect these errors (Fig.~\ref{fig:OriginalClusterState}e). \jc{Multi-qubit Pauli errors can be detected similarly, by considering the syndromes they flip}. Note that $X$ errors on the final cluster state have no effect, \jc{although $X$ errors occurring between two $\CZ$ gates will propagate to $Z$ errors on neighboring qubits}. Errors can be corrected by pairing $(-1)$ syndromes to each other using a minimum weight perfect matching (MWPM) decoder ~\cite{dennis2002topological,edmonds1965paths}, and under this decoder the RHG cluster has a threshold \jc{for local Pauli noise} ~\cite{raussendorf2006fault,raussendorf2007topological}.

This method of creating fault-tolerant cluster states can also be used to generate cluster states that realize the tailored ~\cite{tuckett2018ultrahigh,tuckett2019tailoring,tuckett2020fault} or XZZX ~\cite{ataides2021xzzx} surface codes (Fig.~\ref{fig:SurfaceCodes}b). However, these codes only offer improved thresholds over the usual surface code \jc{when the effective probability of bit-flip errors is suppressed compared to phase-flip errors.} Unfortunately, the 1D teleportation procedure \jc{outlined above} unbiases the noise, converting physical $Z$ errors into logical $X$ errors. We can understand this phenomena in two ways. Firstly, as seen from Eq.~\ref{eq:ZLRewriting} the $Z_{\mathrm{L}}$ operator we use includes physical $X$ operators on qubits $2,4,\dots,2n$ so that $Z$ errors on these qubits anticommute with $Z_{\mathrm{L}}$, and are therefore equivalent to an $X_{\mathrm{L}}$ error. Alternatively, if we consider teleporting the logical information from qubits $1$ and $2$ to, e.g., qubits $7$ and $8$, we see from Eq.~\ref{eq:NewXandZLogicals} that to recover the logical information about the state we require accurate measurements $x_1,\dots,x_6$. A $Z$ error on qubits $2$, $4$, or $6$ would cause us to measure the wrong sign of $x_2$, $x_4$, or $x_6$. This results in the replacement of $Z_{\mathrm{L}}$ with $-Z_{\mathrm{L}}$, which is equivalent to an $X_{\mathrm{L}}$ error.


From either viewpoint, we see that physical $Z$ noise converts to logical $X$ noise. \jc{Consequently, the effective error channel of the 1D teleportation chains, which are subsequently combined to measure the check operators of an error correcting code, is not biased. Thus, the resulting fault-tolerant cluster state will not have an improved threshold even if the measured stabilizers correspond to a code that is specifically designed for biased noise, such as the XZZX surface code. In fact, we find that the XZZX cluster state produced by this approach doesn't perform better than the standard RHG at any bias (at the level of circuit level noise it actually does worse; more details can be found in the supplementary material).}

Naively, one might assume that while the RHG lattice is not robust to $Z$-biased noise, it should be robust to $X$-biased noise, since $X$ errors on the RHG cluster state have no effect on the teleportation. However, even if physical qubits only experience $X$ errors, applying $\CZ_{c,t}$, \jc{where $c$ denotes the control and $t$ the target}, propagates an $X_c$ error to $X_cZ_t$. Thus, if $X$ errors occur during the construction of the RHG lattice, they propagate to $Z$ errors on the final cluster, and we cannot assume we have $X$-biased noise after the construction of the cluster state. By contrast, $\CZ$ gates commute with $Z$ errors, so $Z$-biased noise is compatible with constructing the cluster state.

\subsection{Building the bias-preserving XZZX cluster state}

\begin{figure*}
    \centering
    \includegraphics[width=\textwidth]{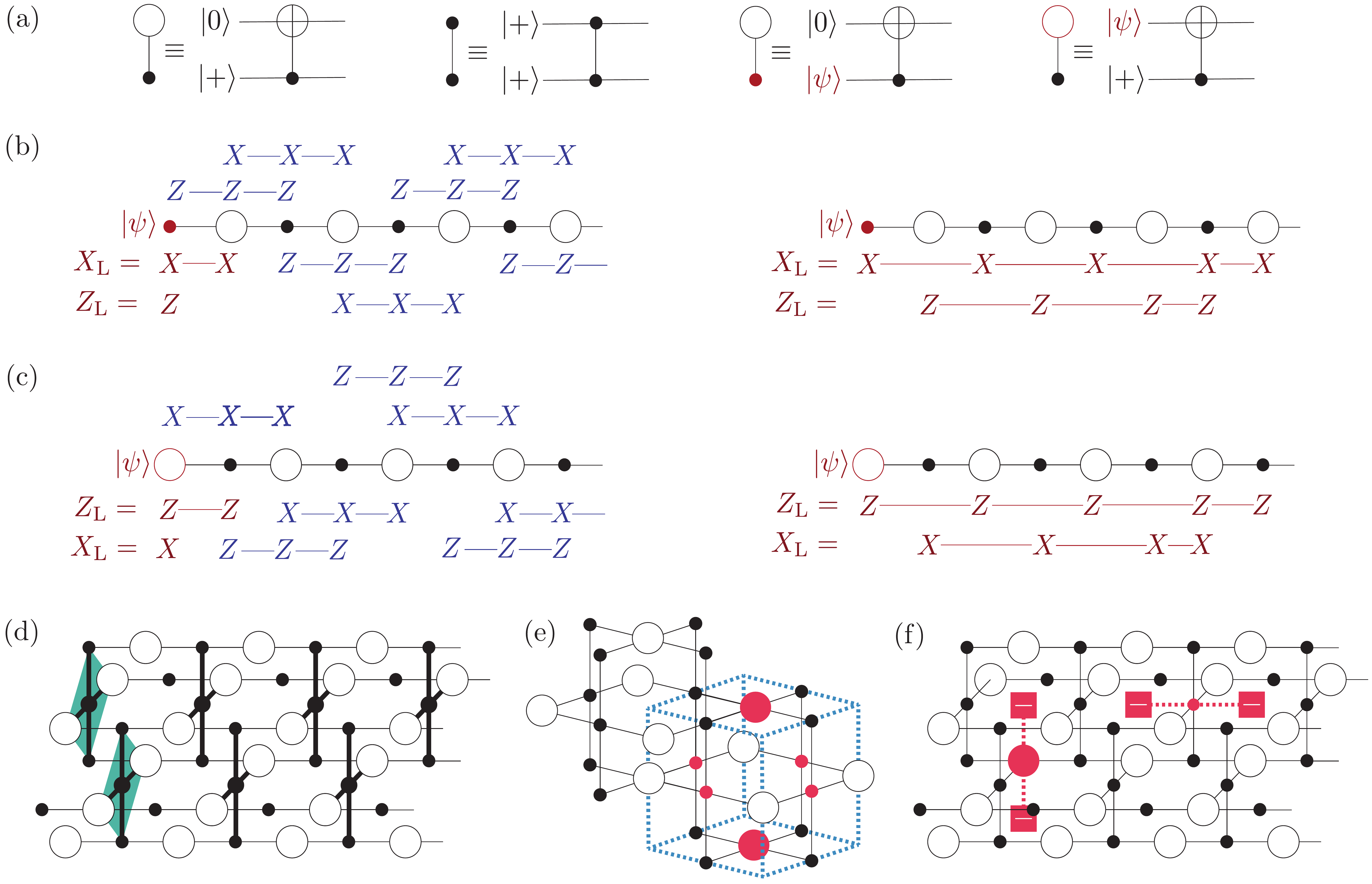}
    \caption{(a) Filled black circles denote $X$-type qubits initialized in the $|+\rangle$ state, open black circles denote $Z$-type qubits initialized in the $|0\rangle$ state, while filled/open red circles denote $X$/$Z$-type qubits initialized in an arbitrary state. Two $X$-type qubits connected by a line have a $\CZ$ gate applied between them, while an $X$-type and $Z$-type qubit connected by a line have a $\CX$ gate applied between them as shown. (b) Left: The logical operators (red) and stabilizers (blue) of the $X$-start 1D cluster state. Right: By multiplying the logical operators by stabilizers, we can rewrite the logical operators to involve only $X$ operators on the first $n$ $X$-type qubits and $Z$ operators on the first $n$ $Z$-type qubits. Measuring the first $n$ $X$-type qubits in the $X$ basis and the first $n$ $Z$-type qubits in the $Z$ basis teleports logical information to qubits $\{2n+1,2n+2\}$. (c) Same, for $Z$-start qubits. (d) By coupling ancilla $X$-type qubits (bold) to an alternating grid of $X$-start and $Z$-start 1D cluster states, we can measure multi-qubit logical operators during teleportation. Here, measuring the $X$ operator on the ancilla qubits measures the XZZX operators of the neighboring qubits. We thus measure the stabilizers of the XZZX code as we teleport. (e) The unit cell of the resulting cluster state, the bias-preserving XZZX cluster state. One can verify that the product of $X$ operators on the highlighted $X$-type qubits and $Z$ operators on the highlighted $Z$-type qubits is a stabilizer of the XZZX cluster state. There is one such stabilizer associated to each cell of the lattice. (f) A $Z$ error on an $X$-type qubit or an $X$ error on a $Z$-type qubit flips the stabilizers of the neighboring cells, allowing us to detect errors. Importantly, $Z$ errors create error chains that are restricted to 2D planes, allowing for improved decoding.}
    \label{fig:GeneralizedClusterState}
\end{figure*}

Based on the discussion above, we see that in order to build a cluster state that realizes the XZZX code with high-threshold, we must start with a 1D teleportation cluster state in which $Z_{\mathrm{L}}$ contains only physical $Z$ operators. To achieve this, we construct our generalized cluster state with two types of qubits, $X$-type and $Z$-type. $X$-type qubits are initialized in $|+\rangle$ and measured in the $X$ basis, as in the usual cluster state, while $Z$-type qubits are initialized in $|0\rangle$ and measured in the $Z$ basis. We'll denote $X$-type qubits by $\newmoon$ and $Z$-type qubits by $\bigcirc$. To entangle neighboring qubits, we will apply different gates depending on the types of qubits we are entangling. To entangle two $X$-type qubits, we apply the usual $\CZ$ gate, while to entangle an $X$- and a $Z$-type qubit, we apply the $\CX$ gate, where the $X$-type qubit is the control qubit and the $Z$-type qubit is the target qubit (Fig.~\ref{fig:GeneralizedClusterState}a). Importantly, the entangling gates are still mutually commuting, and may be applied in any order. \jc{We note that previous cluster states are a subset of our generalized cluster state, in which all qubits are $X$-type.}

Our generalized construction allows us to build two distinct 1D teleportation cluster states, which are illustrated in Figs.~\ref{fig:GeneralizedClusterState}b,c. The first, the $X$-start cluster state, begins with a state $|\psi\rangle$ on the first qubit, \jc{followed by} alternating $Z$-type and $X$-type qubits. While we don't initialize the first qubit in $|+\rangle$, we treat it as an $X$-type qubit during \jc{entanglement} and measurement. The second, the $Z$-start cluster state, begins with a state $|\psi\rangle$ on the first qubit, \jc{followed by} alternating $X$-type and $Z$-type qubits. Here, we treat the first qubit as a $Z$-type qubit during \jc{entanglement} and measurement. 

\jc{After applying the entangling gates, the new logical operators and stabilizers are obtained from the old by conjugating with the entangling gates. This sends
\begin{align}
    Z_i\rightarrow Z_i&\qquad X_i\rightarrow X_i\prod_{\substack{j\in\mathcal{N}_i\\j\in \mathcal{X}}}Z_j\prod_{\substack{k\in\mathcal{N}_i\\k\in \mathcal{Z}}}X_k, & i\in\mathcal{X}\\
    X_i\rightarrow X_i&\qquad Z_i\rightarrow Z_i\prod_{\substack{j\in\mathcal{N}_i\\j\in \mathcal{X}}}Z_j, & i\in\mathcal{Z}
\end{align}
where $\mathcal{X}$ and $\mathcal{Z}$ denote the set of $X$-type and $Z$-type qubits, respectively, and $\mathcal{N}_i$ denotes the neighbors of site $i$. For $X$-start cluster states, we thus have logical operators $X_{\mathrm{L}}=X_1X_2$ and $Z_{\mathrm{L}}=Z_1$ with stabilizers $\{Z_1Z_2Z_3,X_2X_3X_4,\dots\}$, while for $Z$-start we have logical operators $X_{\mathrm{L}}=X_1$ and $Z_{\mathrm{L}}=Z_1Z_2$ with stabilizers $\{X_1X_2X_3,Z_2Z_3Z_4,\dots\}$}. By multiplying the logical operators by stabilizers, we can put them in a form that involves only $X$ operators on the first $n$ $X$-type qubits and only $Z$ operators on the first $n$ $Z$-type qubits. In the case of the $X$-start cluster state, we have
\begin{align}
    X_{\mathrm{L}}&=\left(\prod_{\substack{i\leq 2n\\i\text{ odd}}} X_i\right)X_{2n+1}X_{2n+2}\\
    Z_{\mathrm{L}}&=\left(\prod_{\substack{i\leq 2n\\i\text{ even}}} Z_i\right)Z_{2n+1}
\end{align}
while in the case of the $Z$-start cluster state, we have
\begin{align}
    X_{\mathrm{L}}&=\left(\prod_{\substack{i\leq 2n\\i\text{ even}}} X_i\right)X_{2n+1}\label{eq:XLRewriting2}\\
    Z_{\mathrm{L}}&=\left(\prod_{\substack{i\leq 2n\\i\text{ odd}}} Z_i\right)Z_{2n+1}Z_{2n+2}\label{eq:ZLRewriting2}
\end{align}

The case of \jc{$n=3$} for both clusters is illustrated in Fig.~\ref{fig:GeneralizedClusterState} (b) and (c). This form makes it clear that if we measure the first $n$ $X$-type qubits in the $X$ basis and the first $n$ $Z$-type qubits in the $Z$-basis, the logical operators will be teleported to qubits $\{2n+1,2n+2\}$. \jc{For example, in the case \jc{$n=3$} for the $X$-start cluster state, if we measure qubits \jc{$1-6$} and get outcomes $\{x_1,z_2,x_3,z_4,x_5,z_6\}$ with $x_i,z_i=\pm 1$, Eqs.~\ref{eq:XLRewriting2} and \ref{eq:ZLRewriting2} imply that the logical operators are given by
\begin{equation}
        X_{\mathrm{L}} =x_1x_3x_5X_7X_8,\qquad     Z_{\mathrm{L}} = z_2z_4z_6Z_7.\label{eq:NewXandZLogicals2}
\end{equation}}

Importantly, \jc{in this cluster state} $Z_{\mathrm{L}}$ is the product of physical $Z$ operators and $X_{\mathrm{L}}$ is the product of physical $X$ operators, meaning that physical $Z$ errors cannot cause a logical $X_{\mathrm{L}}$ error. Thus, both teleportation clusters preserve the noise bias. Crucially, these bias-preserving tailored 1D cluster states can be foliated with another stabilizer code like the XZZX code to gain large threshold advantage which was otherwise impossible with the conventional approach.

To build a bias-preserving cluster state realizing the XZZX code, it is most convenient to use an alternating grid of $X$-start and $Z$-start cluster states, as shown in Fig.~\ref{fig:GeneralizedClusterState}d. 
\jc{For each plaquette of the XZZX code, we add an $X$-type ancilla qubit as shown in Fig.~\ref{fig:GeneralizedClusterState}d. These ancillas are also initialized in $|+\rangle$ and entangled with their neighbors following Fig.~\ref{fig:GeneralizedClusterState}a. One can easily verify that measuring an ancilla qubit in the $X$ basis results in measuring the XZZX code stabilizer of the corresponding plaquette during the teleportation.}
The resulting cluster state, the XZZX cluster state, is shown in Fig.~\ref{fig:GeneralizedClusterState}e. The XZZX cluster state can be obtained from the usual RHG cluster state by applying Hadamard ($\mathrm{H}$) gates at the site of $Z$-type qubits, just as the XZZX surface code can be obtained from the usual surface code by conjugating the stabilizers by $\mathrm{H}$ on alternating qubits. However, it is important to physically build the XZZX cluster state with $\CX$ and $\CZ$ gates rather than applying $\mathrm{H}$ gates to the RHG lattice, as an $\mathrm{H}$ operation does not preserve the bias.

Each cell of the XZZX cluster state has four $X$-type qubits and two $Z$-type qubits on its faces, highlighted in Fig.~\ref{fig:GeneralizedClusterState}e; it is straightforward to show that for an XZZX cluster state without errors, the product of the $X$ operators on the $X$-type qubits and the $Z$ operators on the $Z$-type qubits is a stabilizer of the state, so the product of the corresponding $X$ and $Z$ measurements should be $(+1)$. A $Z$ or $Y$ error on an $X$-type qubit flips the syndromes of the neighboring cells, as does an $X$ or $Y$ error on a $Z$-type qubit (Fig.~\ref{fig:GeneralizedClusterState}f). Note that $X$ errors on the overall cluster state have no effect on $X$-type qubits and $Z$ errors on the overall cluster state have no effect on $Z$-type qubits, \jc{although $Z$ errors on $Z$-type qubits occurring between two $\CX$ gates will propagate to $Z$ error on neighboring $X$-type qubits, and similar for $X$ errors on $X$-type qubits. Using a bias-preserving $\CX$ gate ensures that $Z$ errors do not propagate to $X$ or $Y$ errors.} Overall, errors can again be corrected by pairing $(-1)$ syndromes to each other using a MWPM decoder. Importantly, we observe that in the XZZX cluster state $Z$ errors create error chains that are restricted to disconnected 2D planes, while in the RHG cluster state $Z$ errors create error chains that may meander in 3D. Intuitively, this makes decoding the XZZX cluster state easier in the presence of biased noise. This effective reduction in dimensionality of the matching graph in the case biased noise has previously been noted~\cite{brown2020parallelized} as a mechanism for increased thresholds in the tailored~\cite{tuckett2019tailoring} and XZZX~\cite{ataides2021xzzx} surface codes.

\subsection{Comparing the thresholds}

\begin{figure}
    \centering
    \includegraphics[width=\columnwidth]{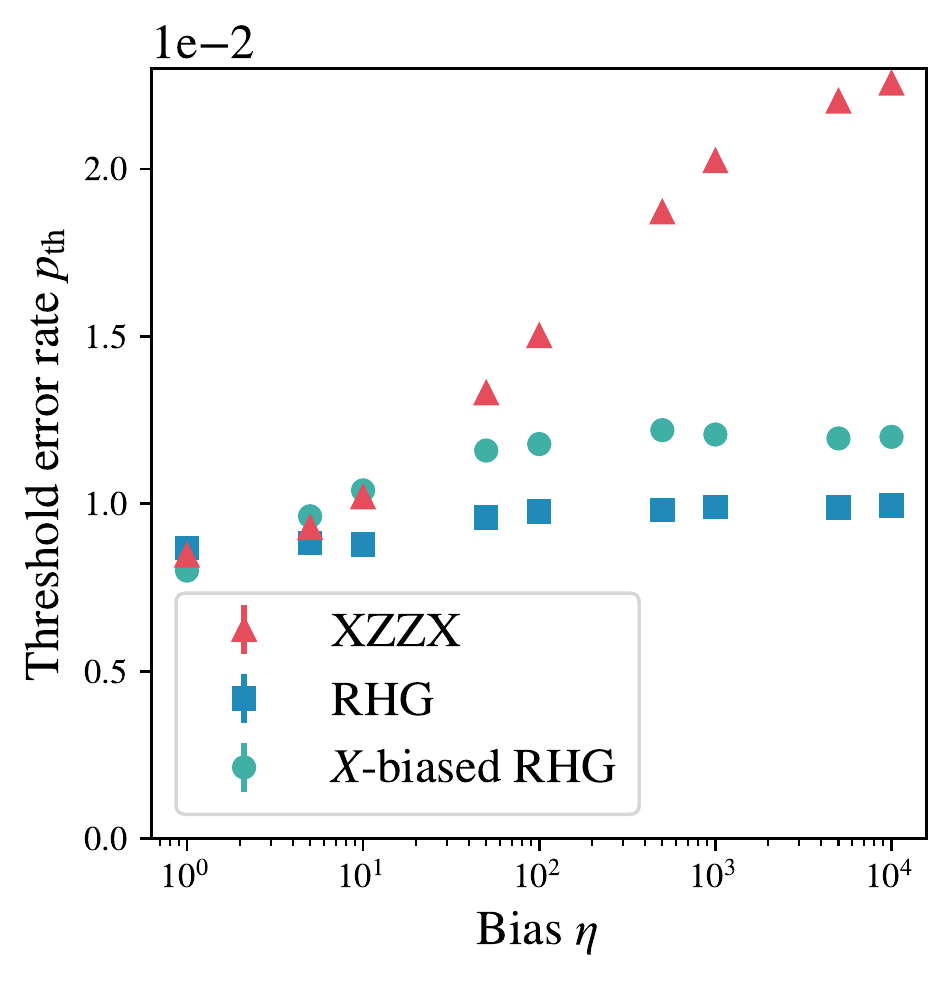}
    \caption{The threshold error rate $p_{\text{th}}$ as a function of bias $\eta$ for the three cluster states we consider, the XZZX and RHG cluster states with $Z$-biased noise and the RHG cluster state with $X$-biased noise. Here $p_{\text{th}}$ refers to the total error rate of the $\CZ$ gate in the corresponding noise model. We see that the XZZX cluster state strongly outperforms both RHG cluster states, although the RHG cluster state has a modestly improved threshold for $X$-biased noise as compared to $Z$-biased. The error bars are smaller than the marker size in the figure.}
    \label{fig:Thresholds}
\end{figure}

To demonstrate the advantage of the XZZX cluster state in the presence of biased noise, we perform full circuit-level noise simulations for both the XZZX cluster state and the usual RHG cluster state. \jc{We do not present results on the cluster obtained by applying the standard foliation approach to the XZZX surface code, as this yields worse thresholds than the RHG state (see the supplemental material).} We use a physically well motivated biased noise model ~\cite{puri2020bias, chamberland2020building,guillaud2021error,darmawan2021practical}, although we do not expect the details of the noise model to \jc{significantly affect our conclusions}. In this model, $\CZ_{c,t}$ gates experience errors $\mathbbm{1}_cZ_t$ and $Z_c\mathbbm{1}_t$ with probability $p_z$, $Z_cZ_t$ with probability $p_z^2$, and all other errors with probability $p_z/\eta$. In addition, $\CX_{c,t}$ gates experience errors $\mathbbm{1}_cZ_t$ and $Z_cZ_t$ with probability $p_z/2$, $Z_c\mathbbm{1}_t$ with probability $p_z$, and all other errors with probability $p_z/\eta$. Finally, during both preparation and measurement, each qubit experiences $Z$ errors with probability $p_z$ and $X$ and $Y$ errors with probability $p_z/\eta$.

In addition, we simulate the RHG cluster state under $X$-biased noise, \jc{to verify our earlier argument that the RHG cluster state should not have a notably higher threshold under $X$-biased noise}. For $X$-biased noise in the RHG cluster state, we use a physically-motivated error model based on a specific implementation of $\CZ$ gates. Note that even when we assume $X$-biased noise on the physical qubits, we do not expect the noise of the $\CZ$ gates to be $X$-biased as $\CZ$ gates do not preserve $X$-bias. In our model, after applying a $\CZ_{c,t}$ gate errors $I_cX_t$, $X_cI_t$, $Z_cX_t$ and $X_cZ_t$ occur with probability $0.375p_x$, $I_cY_t$, $Y_cI_t$, $Z_cY_t$ and $Y_cZ_t$ occur with probability $0.125p_x$, and all other errors occur with probability $p_x/\eta$. In addition to errors during the $\CZ$ gates, during both preparation and measurement $X$ errors occur with probability $p_x$ and $Y$ and $Z$ errors with probability $p_x/\eta$. \jc{We do not simulate the XZZX cluster state with $X$-biased noise, as we expect it to perform worse than with $Z$-biased noise. However, if the physical qubits available are $X$-biased, one can always perform a Pauli frame change to ensure the noise on the XZZX cluster is $Z$-biased.}

In the $Z$-biased noise model, the total error probability of $\CZ$ is $2p_z+p_z^2+12p_z/\eta$ and of $\CX$ gate is $2p_z+12p_z/\eta$. Note that the ratio of the probability of dephasing errors to the probability of errors which cause bit flips is $\eta/6$, so that e.g. $\eta=1000$ corresponds to a ratio of probabilities equal to $166.67$. In the $X$-biased noise model, the total error probability of $\CZ$ is $2p_x+7p_x/\eta$. To compare the cluster states, we will measure the threshold in terms of the error probability of $\CZ$, although we note that the $\CX$ gate in the XZZX cluster state has a near-identical error rate to the $\CZ$ gate for low $p_z$.
For each noise model, we use a MWPM decoder for circuit level-noise to correct the errors ~\cite{dennis2002topological,edmonds1965paths,kolmogorov2009blossom,wang2011surface}. We explain the details of the noise model and our decoder in the Methods section.

We show our results in Fig.~\ref{fig:Thresholds}, where we plot the threshold for biases $1\leq \eta\leq 10000$. We see that at $\eta=1$ the threshold for all three are similar. As we increase the bias, the threshold of the RHG cluster state with $X$-biased noise modestly outperforms the RHG cluster state with $Z$-biased noise; however, as expected, the threshold of our XZZX cluster state strongly outperforms both. For high bias $\eta>1000$, the threshold of the XZZX cluster state has $p_{\text{th}}>2.2\%$, more than doubling the RHG cluster state with $Z$-biased noise which has $p_{\text{th}}<1.0\%$. \jc{Our results also compare favorably to the non-foliated clusters in ~\cite{newman2020generating}. At infinite bias where our noise models become equivalent, the XZZX cluster state moderately outperforms the best non-foliated cluster state, which has a threshold of $~1.93\%$. The non-foliated cluster state requires degree-10 connectivity between qubits, while our cluster state maintains the degree-$4$ connectivity of the usual RHG cluster state. Note that the best cluster state under infinite bias is labelled ``srs" in ~\cite{newman2020generating} and has a threshold of $1.16\%$ under their error model; this translates to a threshold of $1.93\%$ in our error model. We also remark that even if high qubit connectivity is achievable, the thresholds of the non-foliated clusters of ~\cite{nickerson2018measurement,newman2020generating} can likely be improved further using our method to selectively replace some of the $X$-type qubits with $Z$-type qubits.}

\section{Discussion}

We have demonstrated that in the presence of biased noise, the XZZX cluster state offers significant threshold improvements over the usual RHG cluster state. Moreover, this cluster state can be prepared as easily as the standard RHG state. Just like the latter, the entire 3D XZZX cluster state need not be prepared at once and at most two layers of the state need to exist simultaneously at a given time~\cite{raussendorf2007fault}.

While we have focused on the impact of circuit-level Pauli errors in this paper, in some architectures, e.g. photonic, qubit loss is an additional source of error. Fortunately, it is known that the RHG cluster state can simultaneously correct both Pauli errors and qubit loss~\cite{barrett2010fault,whiteside2014upper}. Similarly, in linear optical photonic architectures the entangling gates only succeed probabilistically~\cite{browne2005resource}, but the RHG cluster state is resilient to these errors as well~\cite{auger2018fault}. In both cases, there is a tradeoff between the Pauli noise the threshold can tolerate and the amount of qubit loss or gate failure that can occur; at higher rates of loss or gate failure, the thresholds against Pauli noise are lower. Our XZZX cluster state should be similarly robust to qubit loss and gate failures, and we expect that at a fixed loss and gate failure rate it will have a higher threshold against biased Pauli noise. Quantifying the threshold improvement for realistic values of bias, loss probability, and gate failure rate will be left to future work.

\jc{Leaving aside the question of noise bias for a moment,} the most direct preparation route is to initialize the $Z$-type and $X$-type qubits in $|0\rangle$ and $|+\rangle$ states respectively and then apply the entangling $\CZ$ and $\CX$ gates as described in Fig.~\ref{fig:GeneralizedClusterState}. \jc{This route could be realized in photonic GKP qubits by adapting the methods from ~\cite{bourassa2021blueprint,tzitrin2021fault}.} Alternatively, in linear optical quantum computing (LOQC) architectures with discrete dual-rail encoding, current approaches for RHG state preparation can be adapted for the XZZX cluster state generation~~\cite{li2015resource}. Firstly, photons from single-photon sources can be passed through an interferometric setup consisting of beam-splitters and photon detectors to generate small Greenberger-Horne-Zeilinger (GHZ) states~~\cite{varnava2008good}. Several copies of GHZ states can then be further entangled by destructive measurement of $Z\otimes Z$ and $X\otimes X$ operators using fusion-based Bell measurements~~\cite{browne2005resource,gilchrist2007efficient}. Finally, local Hadamard gates will be required on some of the qubits to prepare the appropriate XZZX cluster state, which can be easily implemented using beam-splitters and phase-shifters in the dual-rail encoding~~\cite{li2015resource,bartolucci2021fusion}. The XZZX cluster state preparation requires fewer Hadamard gates than the preparation of the RHG state.

\begin{figure}
    \centering
    \includegraphics[width=\columnwidth]{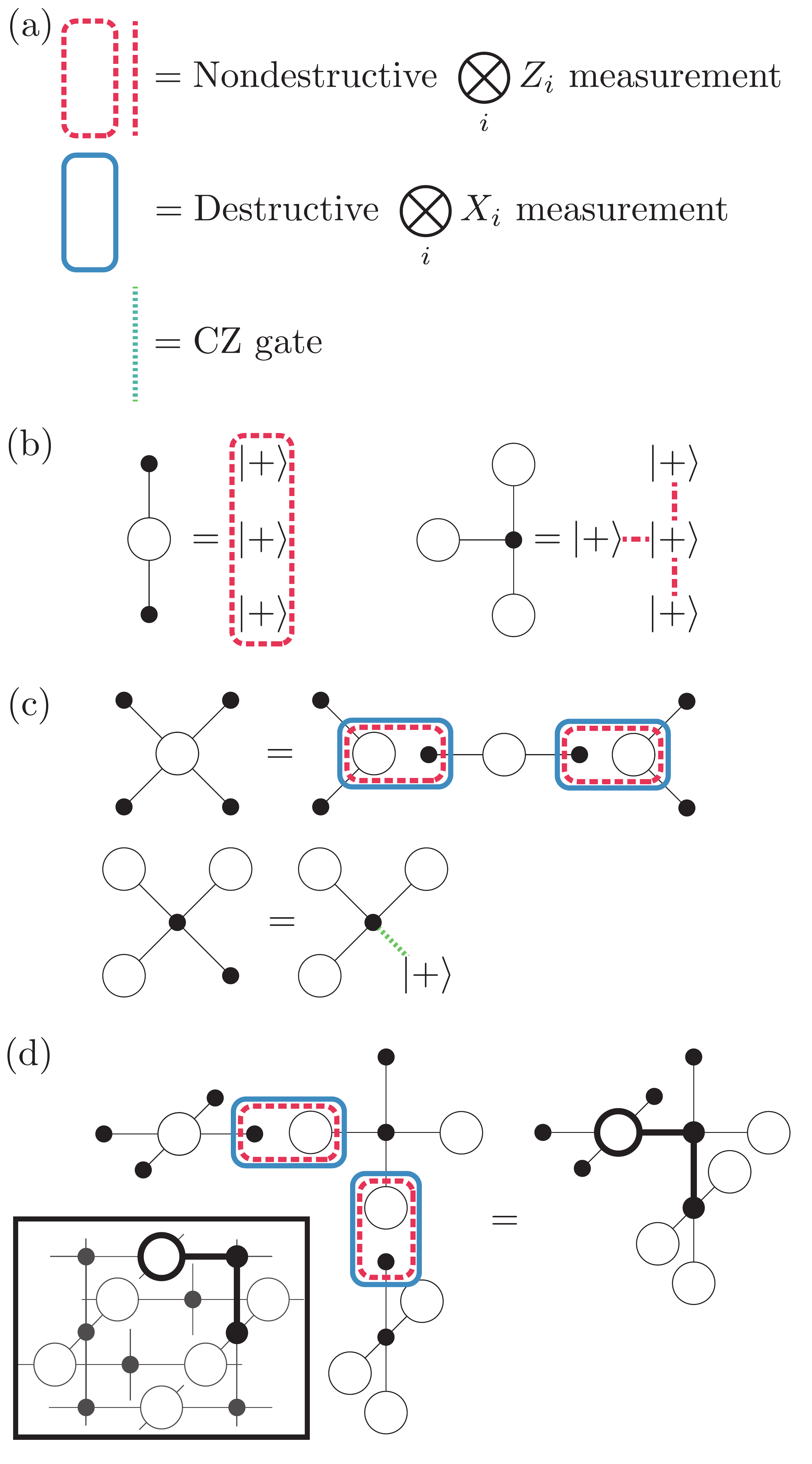}
    \caption{(a) The XZZX cluster state can be built from a product of $|+\rangle$ states using only nondestructive $Z_1\otimes Z_2$ and $Z_1\otimes Z_2\otimes Z_3$ measurements, destructive $X_1\otimes X_2$ measurements, and $\CZ$ gates. \jc{Note that bias-preserving multi-qubit $Z$ measurements can generally be realized nondestructively in biased-noise architectures ~\cite{aliferis2008fault}.} (b) Two elementary cluster states. The left cluster state is constructed by preparing three qubits in the $|+\rangle$ state and performing the three-qubit $Z$ measurement. The right cluster state is constructed by preparing four qubits in the $|+\rangle$ state and performing two-qubit $Z$ measurements as shown. (c) From the elementary cluster states, we can generate larger five-qubit cluster states. The first cluster is formed by fusing three-qubit cluster states using \jc{non-destructive $Z\otimes Z$ followed by destructive $X\otimes X$ measurements.} The second cluster is formed from the four-qubit cluster state by preparing an additional qubit in the $|+\rangle$ state and applying a $\CZ$ gate as shown. (d) The five-qubit clusters can be fused with two-qubit measurements to form the XZZX cluster state. \jc{Here, we illustrate the fusion measurements for the three qubits highlighted in the inset.}}
    \label{fig:Fusions}
\end{figure}

A natural next step is to explore specific qubit architectures that might take advantage of the threshold improvement when the noise becomes biased. Several platforms, such as superconducting fluxonium~~\cite{earnest2018realization}, quantum-dots~~\cite{shulman2012demonstration}, Rydberg atoms~\cite{cong2021hardware}, continuous-variable or bosonic qubits~\cite{grimm2020stabilization,lescanne2020exponential}, and many others, exhibit a biased-noise channel. So far, these qubits have been mostly considered for realizing the circuit model of fault-tolerant topological codes\jc{,} which will require native bias-preserving $\CX$ gates (a bias-preserving gate is one which does not convert a dominant error to a less dominant error and hence does not destroy the noise bias in the underlying hardware). Unlike a diagonal gate such as $\CZ$ which is trivially bias-preserving, a native bias-preserving $\CX$ gate is unphysical in finite-dimensional systems due to a no-go result~\cite{aliferis2008fault,guillaud2019repetition}. Recent works have shown that in some platforms, it is possible to circumvent this no-go result and implement $\CX$ gates in a bias-preserving manner~\cite{puri2020bias,cong2021hardware}. Nonetheless, the schemes proposed so far are experimentally challenging and it is unclear how well they will perform in practice.

On the other hand, there exist standard techniques for implementing high-fidelity, bias-non-preserving $\CX$ gates in a natively biased-noise architecture, for example by combining a $\CZ$ gate with Hadamard gates~\cite{earnest2018realization,darmawan2021practical}.  Such bias-non-preserving $\CX$ gates can then be used to construct a destructive $X\otimes X$ measurement. This measurement is trivially biased because any error in measuring $X\otimes X$ will only lead to a misidentification of the eigenvalue of the $X\otimes X$ operator, which is equivalent to a physical $Z$ error on one of the qubits. Remarkably, the XZZX cluster state can be generated with only bias-preserving $\CZ$ gates, destructive $X\otimes X$ measurements, and nondestructive $Z\otimes Z$ and $Z\otimes Z \otimes Z$ measurements (see Fig.~\ref{fig:Fusions}). Thus, MBQC with the XZZX cluster state may be an avenue for higher-threshold fault-tolerant error correction for biased noise architectures without bias-preserving $\CX$ gates. This opens up a new opportunities to develop high-threshold universal MBQC with a plethora of biased-noise qubits ranging from superconducting qubits to Rydberg atoms.

Finally, while we constructed our high-threshold fault-tolerant cluster state by foliating the XZZX surface code, we have not addressed the possibility of high-threshold non-foliated cluster states. Current versions of non-foliated cluster states are made entirely of $X$-type qubits, which leave every qubit vulnerable to $Z$ errors ~\cite{nickerson2018measurement,newman2020generating}. It is likely that replacing \jc{certain} $X$-type qubits with $Z$-type qubits would lead to higher thresholds, by limiting the qubits where $Z$ errors can affect the cluster. \jc{This could allow for the low-degree cluster states of ~\cite{nickerson2018measurement,newman2020generating} to have a much higher threshold under biased noise, and may further increase the already-high thresholds of their high-degree cluster states under biased noise.} We leave this question open for future exploration.

\

\section{Methods}

\subsection{Circuit level noise model}

For $Z$-biased noise in both the RHG and XZZX cluster state, we use the circuit-level noise model introduced in ~\cite{darmawan2021practical} for biased-noise Kerr cat qubits, although we do not expect the specific error model to noticeably affect our results. We assume the $\CZ_{c,t}$ gates are implemented by evolving under some interaction that is diagonal in the $Z$-basis, e.g. evolving under an interaction of the form $\chi \left[(I_c+Z_t)\mathbbm{1}_t/2+(I_c-Z_c)Z_t/2\right]$ for time $\pi/2\chi$. If we let $p_z$ be the probability of any qubit experiencing a $Z$ error during the operation of this gate, then a $\CZ_{c,t}$ gate experiences $\mathbbm{1}_cZ_t$ and $Z_c\mathbbm{1}_t$ errors with probability $p_z$ and a $Z_cZ_t$ error with probability $p_z^2$. Because the underlying qubits are biased, we assume all other errors occur with probability $p_z/\eta$. A $\CX_{c,t}$ gate experiences $\mathbbm{1}_cZ_t$ and $Z_cZ_t$ errors with probability $p_z/2$, a $Z_c\mathbbm{1}_t$ error with probability $p_z$, and all other errors with probability $p_z/\eta$; justification for this error model can be found in ~\cite{puri2020bias,darmawan2021practical}. Finally, during both preparation and measurement, each qubit experiences $Z$ errors with probability $p_z$ and $X$ and $Y$ errors with probability $p_z/\eta$.

For $X$-biased noise in the RHG cluster state, we use a physically-motivated error model based on a specific implementation of $\CZ$ gates. Let $p_x$ be the probability of an $X$ error occurring on a qubit at any point in the $\CZ$ gate operation, and assume $\CZ_{c,t}$ is implemented by evolving under an interaction of the form $\chi \left[(I_c+Z_c)\mathbbm{1}_t/2+(I_c-Z_c)Z_t/2\right]$ for time $\pi/2\chi$. Then $X$ errors during the evolution result in an error channel whose Pauli-terms are $p_x'I_cX_t\rho I_cX_t+p_x'X_cI_t\rho X_cI_t+p_x'Z_cX_t\rho Z_cX_t+p_x'X_cZ_t\rho X_cZ_t+p_x''I_cY_t\rho I_cY_t+p_x''Y_cI_t\rho Y_cI_t+p_x''Z_cY_t\rho Z_cY_t+p_x''Y_cZ_t\rho Y_cZ_t$, with $p_x':=(2p_x/\pi)\int_0^{\pi/2}\cos^4(\phi)d\phi=0.375p_x$ and $p_x'':=(2p_x/\pi)\int_0^{\pi/2}\cos^2\sin^2(\phi)d\phi=0.125p_x$. We assume every other Pauli error occurs with probability $p_x/\eta$. Note that even if the qubits only experience $X$ errors, evolving under the $\CZ$ interaction converts $X$ errors into $Z$ errors. Indeed, it does not appear that an $X$-bias-preserving $\CZ$ gate is physical. In addition to errors during the $\CZ$ gates, during both preparation and measurement $X$ errors occur with probability $p_x$ and $Y$ and $Z$ errors with probability $p_x/\eta$.

\begin{figure}
    \centering
    \includegraphics[width=\columnwidth]{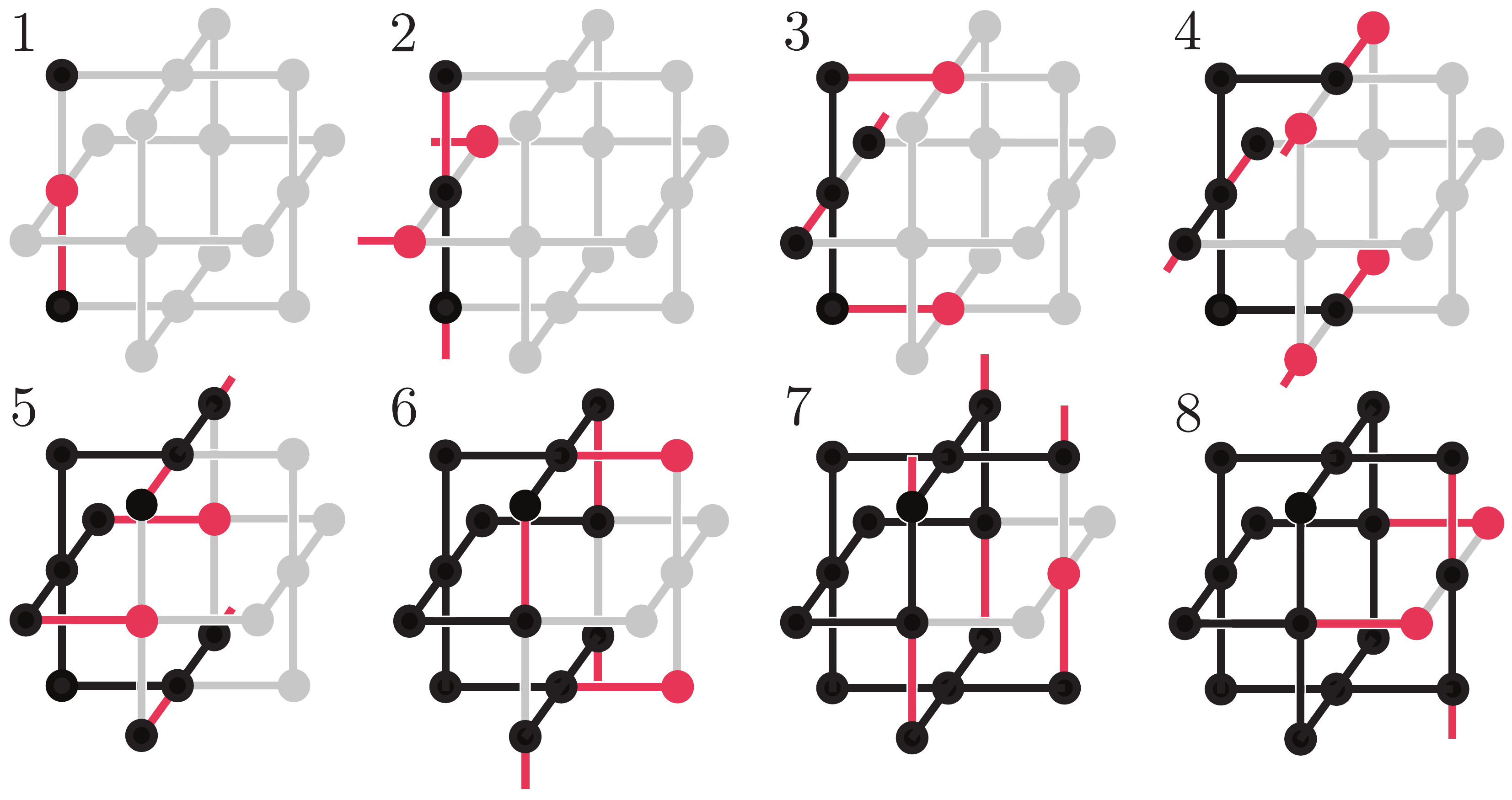}
    \caption{The order in which we apply our entangling gates to build the both the RHG and XZZX cluster state. Red edges are gates that are performed at that timestep, red vertices are qubits that are initialized for that timestep, and black edges/vertices are gates/qubits that have already been introduced in a previous timestep. We construct our cluster state layer-by-layer, so our gate order is compatible with only a 2D layer of the cluster state existing at any point in time. Note that \jc{timesteps $7$ and $8$} coincide with \jc{timesteps $1$ and $2$} for the next layer. Our choice of gate order ensures that each qubit can be initialized, entangled with its four neighbors, and then measured without ever being idle.}
    \label{fig:GateOrder}
\end{figure}

To build the cluster state, we apply our entangling gates in the order shown in Fig.~\ref{fig:GateOrder} for both the RHG and XZZX cluster states, although we expect other orderings to perform similarly. Our ordering is compatible with building the cluster state layer-by-layer in time, as we can measure the leftmost qubits as soon as they are entangled with their neighbors without waiting for the rightmost qubits to be initialized. Our ordering also ensures that qubits can be initialized, entangled, and measured without ever sitting idle.

\subsection{The MWPM decoder}

To match error syndromes in a circuit-level noise model, we use the approach originally developed in Ref. ~\cite{wang2011surface}. Given error syndrome locations $\{e_1,...,e_n\}$, we want to determine the most likely pairing of the error syndromes. This can be done via a MWPM algorithm, which takes as input the weights $W_{e_i,e_j}$ between all pairs of error syndromes, and outputs a matching that minimizes the weights between them ~\cite{edmonds1965paths,kolmogorov2009blossom}.

\begin{figure}
    \centering
    \includegraphics[width=\columnwidth]{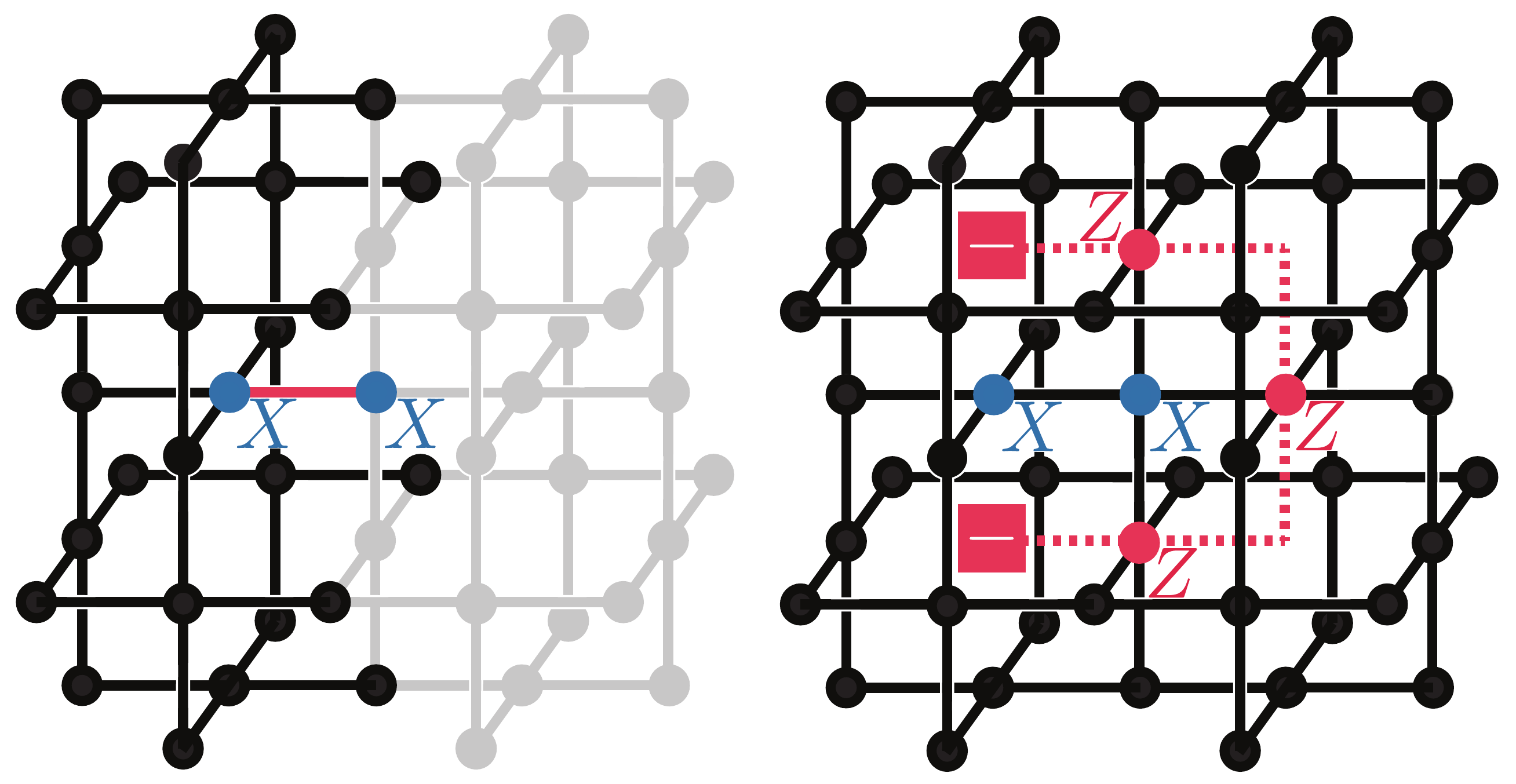}
    \caption{An example of propagating errors in the RHG cluster state. If an $X\otimes X$ error occurs after the CZ gate highlighted in red while building the RHG state, subsequent CZ gates spread $X$ errors to $Z$ errors on neighboring qubits. The $X\otimes X$ error on the CZ gate results in a Pauli error on the final state that involves three $Z$ operators and two $X$ operators. When we measure the qubits in the $X$ basis, the $X$ errors do nothing, while the $Z$ errors result in erroneous measurements on the red highlighted qubits. Overall, the $X\otimes X$ error results in the error syndrome shown.}
    \label{fig:ErrorsToWeights}
\end{figure}

We determine the weights between two syndromes $W_{e_i,e_j}$ by approximating $-\log(P_{e_i,e_j})$, where $P_{e_i,e_j}$ is the probability that $e_i$ and $e_j$ were connected by a string of errors. We generate $W_{e_i,e_j}$ by iterating through all possible errors. For each possible preparation, measurement, or gate error during the creation of the cluster state, we determine the effect of that error on the syndromes. We illustrate a representative case in Fig.~\ref{fig:ErrorsToWeights}. We consider only errors that result in exactly two error syndromes. Each time two locations $s_1$ and $s_2$ are connected by such an error, we increment the probability $p_{s_1s_2}$ associated to that connection by the probability of that error. We then define the local weight between two syndrome locations to be $w_{s_1s_2}=-\log(p_{s_1s_2})$, with $w_{s_1s_2}=\infty$ if $s_1$ and $s_2$ are not connected by any error. The true weight $W_{e_1e_2}$ between two syndrome locations $e_i$ and $e_j$ is then defined to be the smallest sum of local weights for all paths connecting $e_i$ and $e_j$:
\begin{equation}
    W_{e_ie_j}=\min_{\{s\}}\left[w_{e_is_1}+w_{s_1s_2}+\cdots+ w_{s_me_j}\right].
\end{equation}
The weights $W$ can be efficiently determined from the local weights $w$ by standard pathfinding algorithms ~\cite{dijkstra1959note}.

\subsection{Determining the threshold}

To determine the threshold, we use the method of ~\cite{wang2003confinement}. For a given $\eta$, we simulate a cluster state of size $d$ experiencing errors at a physical error rate $p_z$ and determine the logical error probability $p_L$; we repeat this procedure for various values of $d$ and $p_z$. Near threshold $p_{\text{th}}$ and for large values $d$, the logical error rate has the scaling form
\begin{equation}
 p_L = f\left((p-p_{\text{th}})d^{1/\nu}\right)   
\end{equation}
for some scaling parameter $\nu$. Expanding $f(x)\approx A+Bx+Cx^2$ for small $x$, we fit the values of $A$, $B$, $C$, $\nu$, and $p_{\text{th}}$ to give an estimate of $p_{\text{th}}$.

The error bars in our estimate for $p_{\text{th}}$ come from the uncertainty in this fit. The uncertainty originates from statistical uncertainty in our estimate for $p_L$, but does not include possible systematic errors from our approximation to $f(x)$ or finite-size effects, as is standard when using the method of ~\cite{wang2003confinement}.

\jc{For each threshold estimate, we take data at four different distances $d$ and $12-15$ different values of $p_{\textrm{CZ}}$ within $20\%$ of our threshold value. For each value of $d$ and $p_{\CZ}$ we estimate the value of $p_L$ through some number $n_S$ of Monte Carlo samples, which ranges from $n_S=68,000$ for our smallest systems to $n_S=1600$ for our largest system. When performing the fits, we weight the data according to the statistical uncertainty, so that data points with less uncertainty are weighted more highly in the fit. We include more details of the distances $d$ and the values $p_{\CZ}$ we consider in the supplemental material}

\

\begin{acknowledgments}
 The authors thank Steven Touzard, Guillaume Dauphinais, Mahnaz Jafarzadeh, and Ilan Tzitrin for discussions on MBQC and photonic quantum computing, as well as Benjamin J. Brown for discussions on fault-tolerant cluster states. We thank Benjamin J. Brown, Priya Nadkarni, and Kenneth R. Brown for a critical reading of the manuscript. This material is based on work supported by the National Science Foundation (NSF) under Award No. 2137740. Any opinions, findings, and conclusions or recommendations expressed in this publication are those of the authors and do not necessarily reflect the views of NSF.
\end{acknowledgments}

\bibliography{thebibliography.bib}

\end{document}


\title{Supplemental Material: Tailored cluster states with high threshold under biased noise}

\author{Jahan Claes}
\affiliation{Department of Applied Physics, Yale University, New Haven, Connecticut 06511, USA}
\affiliation{Yale Quantum Institute, Yale University, New Haven, Connecticut 06511, USA}
\author{J. Eli Bourassa}
\affiliation{Xanadu, Toronto, Ontario, M5G 2C8, Canada}
\author{Shruti Puri}
\affiliation{Department of Applied Physics, Yale University, New Haven, Connecticut 06511, USA}
\affiliation{Yale Quantum Institute, Yale University, New Haven, Connecticut 06511, USA}

\maketitle

\renewcommand{\theequation}{S\arabic{equation}}
\renewcommand{\thefigure}{S\arabic{figure}}
\renewcommand{\thetable}{S\arabic{table}}
\renewcommand{\bibnumfmt}[1]{[S#1]}
\renewcommand{\citenumfont}[1]{S#1}
\renewcommand{\thesection}{S\Roman{section}}

In this supplement, we provide additional details on several aspects mentioned in the main paper. First, we derive the stabilizers of the RHG and XZZX cluster states. Next, we demonstrate two ways of foliating the XZZX surface code using only $X$-type qubits and show that this doesn't lead to any better thresholds than the usual RHG lattice. We also give more details on the thresholds for the circuit-level noise model, including an example of the data we use to find the threshold. Finally, we include simulations of a simpler phenomenological noise model, which may be useful for comparison to other works which also use phenomenological noise models, such as~\cite{nickerson2018measurement,newman2020generating}.

\section{Stabilizers of the RHG and XZZX cluster states}

To derive the final stabilizers of a cluster state, we track how the stabilizers change during the construction of the cluster state. To begin, we initialize all $X$-type qubits in $|+\rangle$ and all $Z$-type qubits in $|0\rangle$, so the stabilizers are
\begin{equation}
\begin{split}
        X_i, &\qquad i\in\mathcal{X}\\
    Z_i, &\qquad i\in\mathcal{Z}.
\end{split}
\end{equation}
We want to apply entangling gates to this state; these entangling gates will change our stabilizers. To understand how the stabilizers will change, consider a state $|\phi_0\rangle$ with stabilizer $S$, so that $S|\phi_0\rangle=|\phi_0\rangle.$. If we consider the state $|\phi_1\rangle = U|\phi_0\rangle$ for some operator $U$, it is straightforward to check that $USU^\dagger|\phi_1\rangle=|\phi_1\rangle$. Therefore, applying an operator $U$ to a state sends $S\rightarrow USU^\dagger$.

To find the stabilizers after we apply the entangling gates $\CZ_{c,t}$ and $\CX_{c,t}$, we simply need to conjugate our original $X_i$ and $Z_i$ operators by the entangling gates. We make use of the circuit identities
\begin{equation}\begin{split}
    \CZ_{c,t} X_c \CZ_{c,t}^\dagger &= X_c Z_t\\
    \CZ_{c,t} X_t \CZ_{c,t}^\dagger &= Z_c X_t\\
    \CX_{c,t} X_c \CX_{c,t}^\dagger &= X_c X_t\\
    \CX_{c,t} Z_t \CX_{c,t}^\dagger &= Z_c Z_t
\end{split}\end{equation}
and note that all other $X$ and $Z$ operators are unchanged under conjugation. Conjugating our $X$ and $Z$ stabilizers by the entangling gates then gives
\begin{equation}
    \begin{split}
    X_i\prod_{\substack{j\in\mathcal{N}_i\\j\in \mathcal{X}}}Z_j\prod_{\substack{k\in\mathcal{N}_i\\k\in \mathcal{Z}}}X_k, &\qquad i\in\mathcal{X}\\
    Z_i\prod_{\substack{j\in\mathcal{N}_i\\j\in \mathcal{X}}}Z_j, & \qquad i\in\mathcal{Z}.\label{eq:stabilizers}
    \end{split}
\end{equation}
We see that conjugating by the entangling gates spreads the support of the stabilizers to the neighboring qubits.

\begin{figure}
    \centering
    \includegraphics[width=\columnwidth]{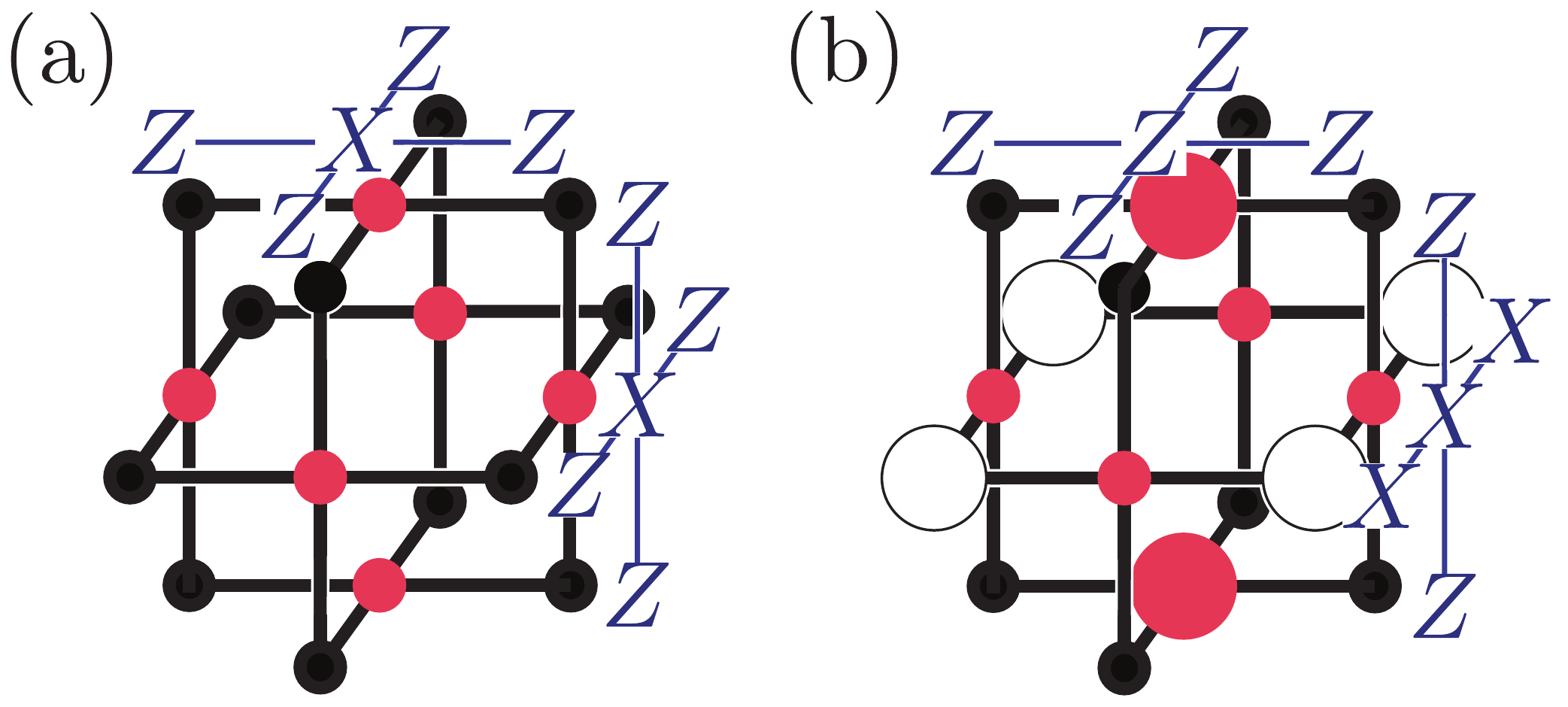}
    \caption{Two stabilizers of the (a) RHG and (b) XZZX cluster states, centered on two of the highlighted qubits. If we multiply all stabilizers centered on the highlighted qubits together we get a new stabilizer with support only on the highlighted qubits.}
    \label{fig:FindingStabilizers}
\end{figure}

As an example, we can apply Eq. ~\ref{eq:stabilizers} to the unit cells of the RHG and XZZX cluster states. We illustrate two stabilizers of the RHG cluster state in Fig. ~\ref{fig:FindingStabilizers}a, where the center of each stabilizers is at a qubit $i$ highlighted in red. Multiplying the two stabilizers shown, along with the four other stabilizers centered at the other four highlighted qubits, results in a new stabilizer that is simply the product of the $X$ operators on the highlighted qubits. This stabilizer is what allows the RHG cluster state to detect errors, as demonstrated in the main text.

As a second example, we illustrate two stabilizers of the XZZX cluster state in Fig. ~\ref{fig:FindingStabilizers}b, where again the center of each stabilizers is at a qubit $i$ highlighted in red. Multiplying the two stabilizers shown, along with the four other stabilizers centered at the other four highlighted qubits, results in a new stabilizer that is the product of the $X$ operators on the highlighted $X$-type qubits and the $Z$ operators on the highlighted $Z$-type qubits. This stabilizer then allows the XZZX cluster state to detect errors as well.

\section{Foliating the XZZX surface code without the generalized cluster state}

Instead of using our generalized cluster state, one could attempt to foliate the XZZX surface code using previous methods. In the original formulation, foliation was only defined for CSS codes, in which stabilizers are either products of $X$ operators or products of $Z$ operators~\cite{bolt2016foliated}. More recent work has extended foliation to arbitrary stabilizer codes~\cite{brown2020universal}. However, it is not always possible to measure the stabilizers of a non-CSS code by coupling ancilla qubits to the 1D teleportation chain. Instead, one may have to also couple the ancilla qubits to each other in order to properly measure the stabilizers (see~\cite[Sec. IV.E]{brown2020universal}).

\begin{figure}
    \centering
    \includegraphics[width=\columnwidth]{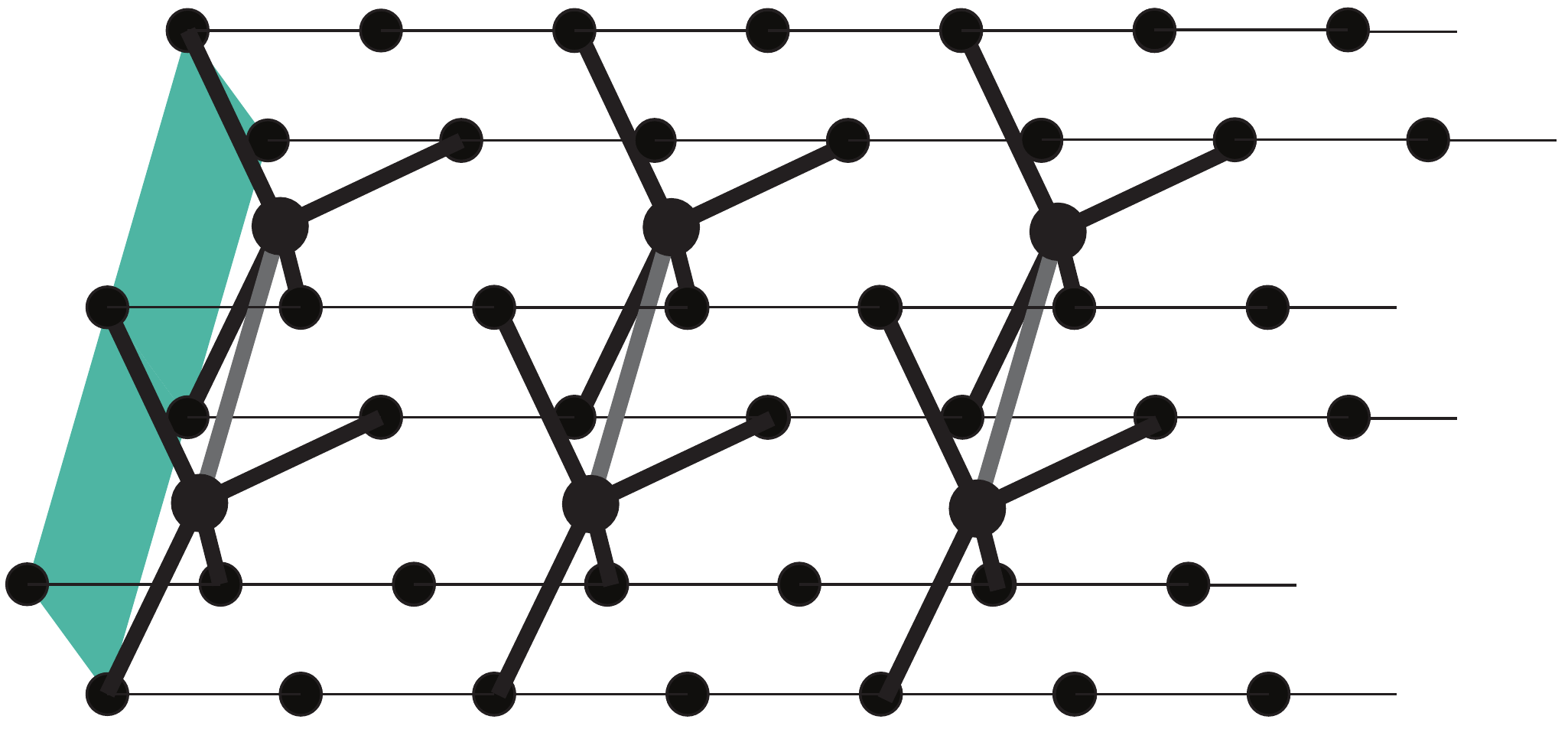}
    \caption{Foliating the XZZX surface code with the method of~\cite{brown2020universal}. For each plaquette of the XZZX surface code, we add an ancilla qubit whose measurement gives the value of the corresponding stabilizer. In order for us to be able to simultaneously measure all the stabilizers of the XZZX cluster state, we also have to couple neigboring ancilla qubits; this additional coupling is shown in grey. For more details on the subtleties of simultaneously measuring non-CSS stabilizers, we refer to~\cite[Sec. IV.E]{brown2020universal}.}
    \label{fig:XZZXBadFoliation}
\end{figure}

We show how to foliate the XZZX surface code using the foliation method of~\cite{brown2020universal} in Fig. ~\ref{fig:XZZXBadFoliation}. For each plaquette of the XZZX surface code, we include an ancilla qubit whose measurement gives the value of the corresponding stabilizer. In order for us to simultaneously measure all the stabilizers of the XZZX code, it is necessary to couple the neighboring ancilla qubits together. This additional coupling is shown in grey. Note that this increases the degree of the cluster state; while the teleportation qubits still have degree 4, the ancilla qubits have degree 8.

We do not include results on this foliated version of the XZZX surface code in the main text, as we do not expect it to be competitive with either the RHG or XZZX cluster state. First, it requires a greater number of $\CZ$ gates to construct which leads to a higher level of noise per qubit, yet its decoding graph is identical to that of the usual RHG lattice, so this increase in noise will not be compensated by more effective decoding. Second, since it is built out of non-bias-preserving 1D teleportation chains, we do not expect the XZZX code to offer an advantage. Indeed, we simulated this model and found circuit-level noise thresholds at $\eta=1,100,10000$ of roughly $.45\%$, $.75\%$, and $.8\%$, respectively. These thresholds are clearly lower than the corresponding thresholds for the usual RHG cluster state and much lower than the corresponding thresholds of the XZZX cluster state.

\section{The XZZX and RHG cluster states with circuit-level noise}

\begin{figure*}
    \centering
    \includegraphics[width=2\columnwidth]{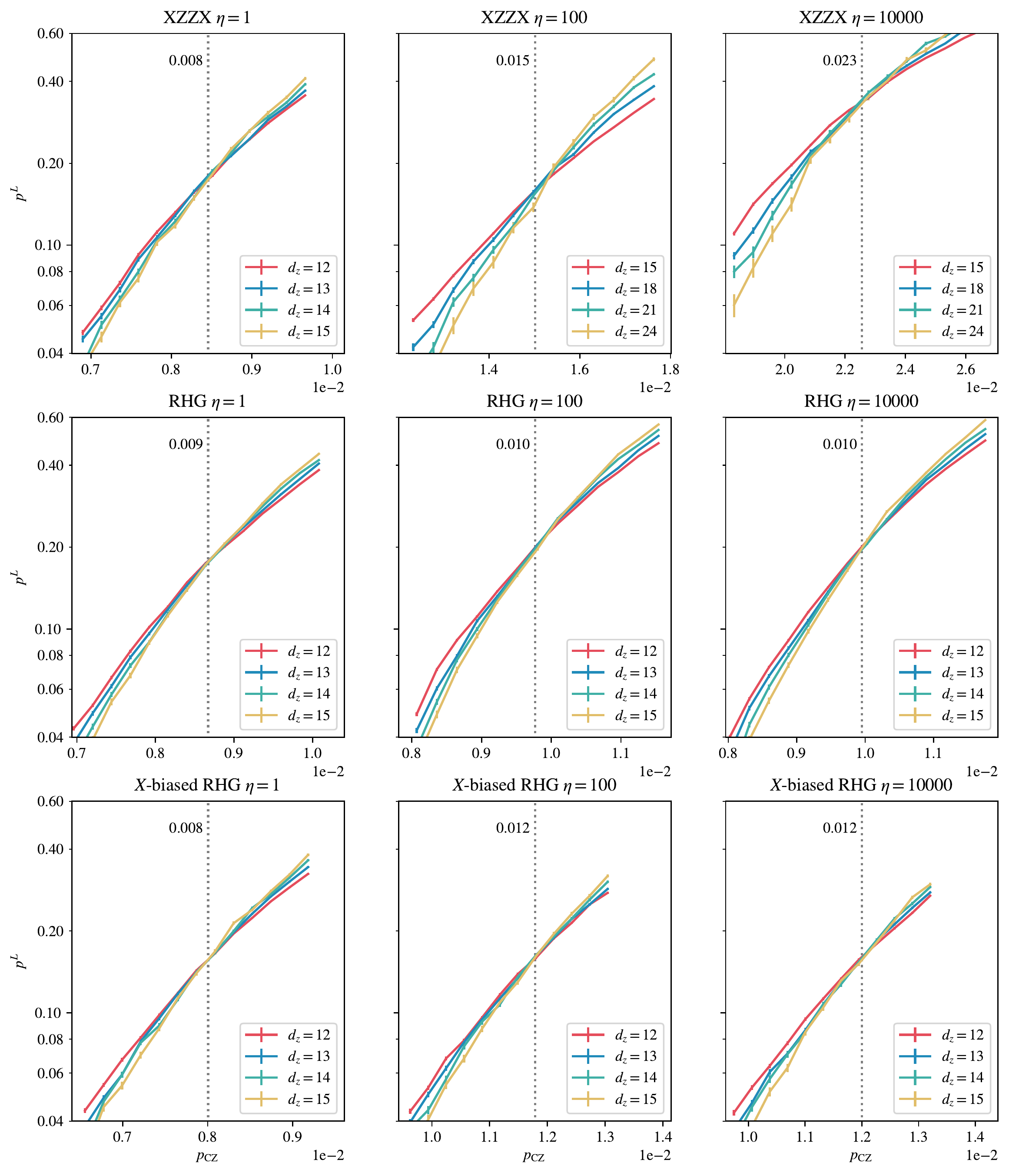}
    \caption{Threshold plots for the three circuit-level noise simulations at $\eta=1,100,10000$. The $x$-axis is the error rate on the $\CZ$ gate, the $y$ axis is the logical error rate after error correction. We plot curves for various system sizes. The intersection point of these curves gives the threshold error rate $p_{\text{th}}$.}
    \label{fig:CircuitLevelNoiseThresholdsExample}
\end{figure*}

To give an example of our threshold calculations and the sub-threshold scaling of our cluster states, we include threshold plots for $\eta=1,100,10000$ in Fig. ~\ref{fig:CircuitLevelNoiseThresholdsExample}. In these plots, the $x$-axis is the total error rate of the $\CZ$ gate, and the $y$-axis is the total logical error rate after error correction. We simulate the cluster state for different sizes $d_z$. The cluster state has dimensions $d_z\times d_z\times d_z$ for the RHG cluster states as well as the XZZX cluster state for $\eta<100$, and it has dimensions $(d_z/3)\times d_z\times d_z$ for the XZZX cluster state for $\eta\geq 100$, with the shorter dimension associated with the logical $X$ operator. The intersection point of these curves gives the threshold logical error rate $p_{\text{th}}$ for that model.

By fitting the data to the scaling function~\cite{wang2003confinement}
\begin{equation}
    \begin{split}
         p_L &= f\left((p-p_{\text{th}})d^{1/\nu}\right)\\
        f(x)&=A+Bx+Cx^2
    \end{split}\label{eq:FittingFunction}
\end{equation}
as described in the main text, and taking the length scale $d$ to be $d_z$, we extract the threshold $p_{\text{th}}$, which is what is reported in Fig. 4 of the main text.

\section{The XZZX and RHG cluster states with phenomenological noise}

\begin{figure*}
    \centering
    \includegraphics[width=2\columnwidth]{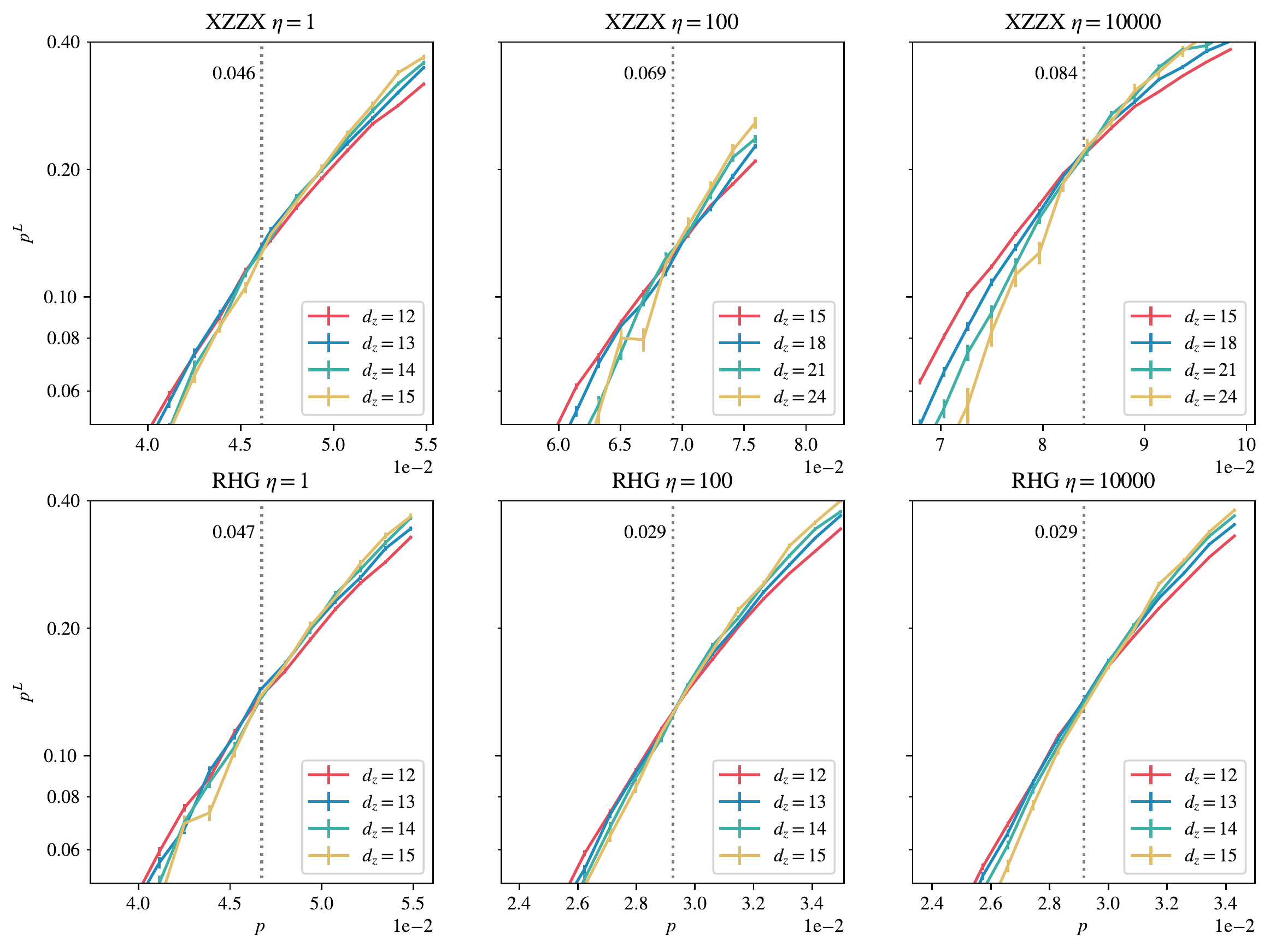}
    \caption{Threshold plots for the two phenomenological noise simulations at $\eta=1,100,10000$. The $x$-axis is the total phenomenological error rate $(p_z+2p_z/\eta)$, the $y$ axis is the logical error rate after error correction. We plot curves for various system sizes. The intersection point of these curves gives the threshold error rate $p_{\text{th}}$.}
    \label{fig:PhenomNoiseThresholdsExample}
\end{figure*}

\begin{figure}
    \centering
    \includegraphics[width=\columnwidth]{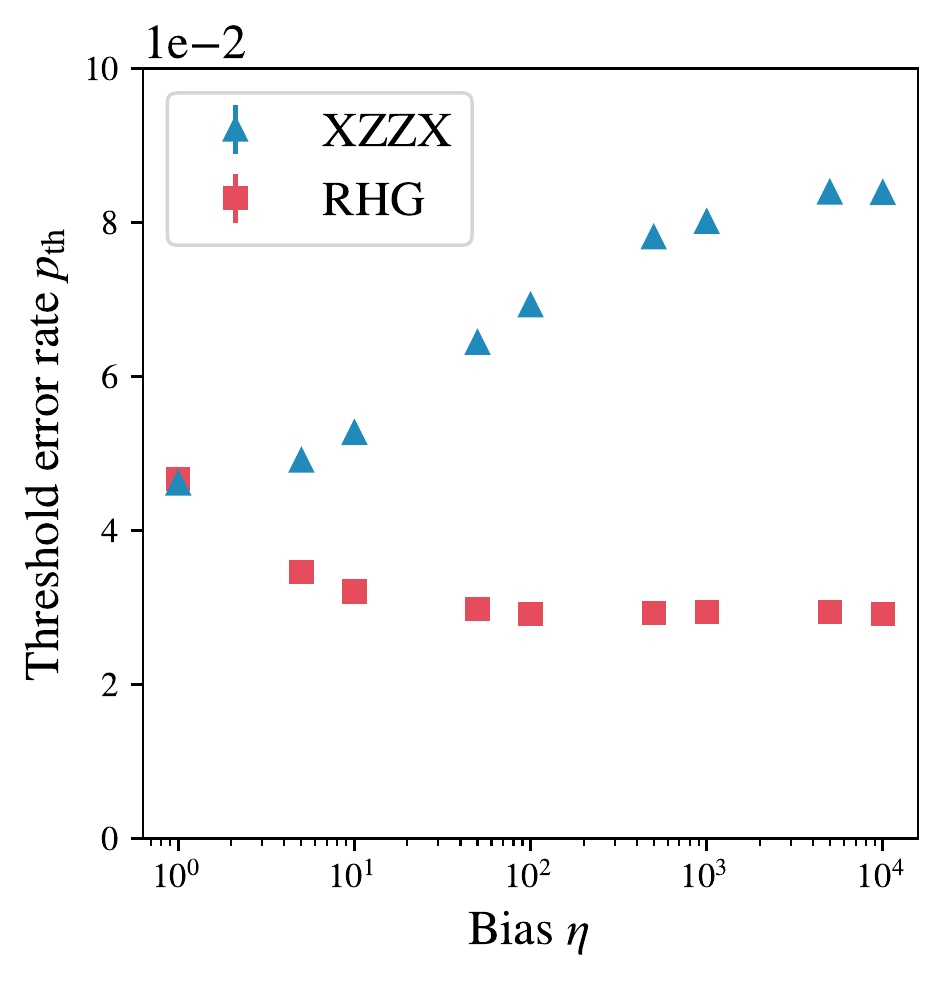}
    \caption{The threshold error rate $p_{\text{th}}$ as a function of bias $\eta$ for the two cluster states, the XZZX and RHG cluster states with phenomenological $Z$-biased noise. Here $p_{\text{th}}$ refers to the total phenomenological error rate, $(p_z+2p_z/\eta)$. We see that the XZZX cluster state strongly outperforms RHG cluster state. In this figure, the error bars are smaller than the marker size.}
    \label{fig:PhenomNoiseThresholdsGraph}
\end{figure}

To simplify calculations and more easily compare the XZZX cluster state to other results in the literature, we can evaluate the thresholds using a phenomenological noise model for the cluster state. In this noise model, we assume that uncorrelated errors occur on each qubit just before the qubit is measured (or equivalently, just after each qubit is initialized). We assume $Z$ errors occur with probability $p_z$, and $X$ and $Y$ errors occur with probability $p_z/\eta$. We do not consider the RHG cluster state with $X$-biased phenomenological noise, because as noted in the main text $X$-biased noise on the final RHG cluster state is not compatible with building the RHG cluster state with $\CZ$ gates.

In Fig. ~\ref{fig:PhenomNoiseThresholdsExample} we show threshold plots for $\eta=1,100,10000$. Here, the $x$-axis is the total error rate on each qubit, $(p_z+2p_z/\eta)$, and the $y$-axis is the total logical error rate after error correction. We again simulate the cluster state for different sizes $d_z$, with the same dimensions as for our circuit-level noise model. The intersection point of these curves gives the threshold logical error rate $p_{\text{th}}$ for that model.

To determine $p_{\text{th}}$, we again fit to Eq. ~\ref{eq:FittingFunction}. Our threshold error rates are plotted in Fig. ~\ref{fig:PhenomNoiseThresholdsGraph}.

\bibliography{thebibliography.bib}